\pdfoutput=1
\documentclass[a4paper,11pt]{article}

\usepackage[margin=1in,includefoot]{geometry}
\usepackage[utf8]{inputenc}
\usepackage[T1]{fontenc}
\usepackage{lmodern,microtype,mathtools,amsthm,amssymb}
\usepackage[hidelinks]{hyperref}
\usepackage{graphicx,xcolor,tikz}

\hypersetup{
  pdftitle={Distinguishing classes of intersection graphs of homothets or similarities of two convex disks},
  pdfauthor={Mikkel Abrahamsen, Bartosz Walczak},
  colorlinks,
  linkcolor={red!50!black},
  citecolor={blue!50!black},
  urlcolor={blue!80!black}
}

\title{Distinguishing Classes of Intersection Graphs of Homothets~or~Similarities of Two Convex Disks\thanks{To appear in \emph{The Electronic Journal of Combinatorics}.
A short version of this paper was presented at the 39th International Symposium on Computational Geometry (SoCG 2023)~\cite{abrahamsen2023distinguishing}.}}
\date{}

\author{Mikkel Abrahamsen\thanks{Basic Algorithms Research Copenhagen (BARC), Department of Computer Science, University of Copenhagen, Denmark.
The author was partially supported by the Independent Research Fund Denmark, grant 1054-00032B, and by the Carlsberg Foundation, grant CF24-1929.}\and Bartosz Walczak\thanks{Department of Theoretical Computer Science, Faculty of Mathematics and Computer Science, Jagiellonian University, Kraków, Poland.
The author was partially supported by the National Science Center of Poland, grant 2015/17/D/ST1/00585.}}

\DeclarePairedDelimiter{\set}{\{}{\}}
\DeclarePairedDelimiter{\abs}{\lvert}{\rvert}
\DeclarePairedDelimiter{\norm}{\lVert}{\rVert}
\DeclareMathOperator{\dist}{dist}
\DeclareMathOperator{\diam}{diam}
\DeclareMathOperator{\area}{area}
\let\int\undefined
\DeclareMathOperator{\int}{int}
\let\sim\undefined
\DeclareMathOperator{\sim}{sim}

\newcommand{\setR}{\mathbb{R}}
\newcommand{\setN}{\mathbb{N}}
\newcommand{\calF}{\mathcal{F}}
\newcommand{\calH}{\mathcal{H}}
\newcommand{\calX}{\mathcal{X}}
\newcommand{\Ghom}{G^{\mathrm{hom}}}
\newcommand{\Gsim}{G^{\mathrm{sim}}}
\newcommand{\Gaff}{G^{\mathrm{aff}}}
\newcommand{\origin}{\mathbf{0}}
\newcommand{\rots}{\mathcal{R}}
\newcommand{\rep}{\mu}
\newcommand{\intrep}{\bar\mu}
\newcommand{\refl}[1]{#1^{\mathrm{r}}}
\newcommand{\simrefl}{\refl{\sim}}
\newcommand{\disk}[2]{D_{#2}(#1)}

\let\leq\leqslant
\let\geq\geqslant
\let\setminus\smallsetminus
\let\epsilon\varepsilon

\newtheorem{theorem}{Theorem}
\newtheorem{lemma}[theorem]{Lemma}

\theoremstyle{definition}

\newtheorem{construction}[theorem]{Construction}

\makeatletter
\def\th@subclaim{\thm@headfont{\itshape}\normalfont}
\makeatother

\theoremstyle{subclaim}
\newtheorem{subclaim}{Claim}
\numberwithin{subclaim}{theorem}

\makeatletter
\renewenvironment{proof}[1][\proofname]{%
  \par\pushQED{\qed}\normalfont\trivlist
  \item[\hskip\labelsep\itshape #1\@addpunct{.}]\ignorespaces}{%
  \popQED\endtrivlist\@endpefalse}
\makeatother
\newcommand{\opentriangle}{%
  \leavevmode\hbox to .77778em{\hfil\tikz[x=.6em,y=.625em,line width=.4pt]{
    \draw (0,.5) -- (1,1) -- (1,0) -- cycle;
  }\hfil}}

\newenvironment{claimproof}{%
  \begin{proof}}{%
  \end{proof}}

\makeatletter
\newdimen\new@belowdisplayshortskip
\new@belowdisplayshortskip\abovedisplayskip
\divide\new@belowdisplayshortskip 3
\def\shortskipafterdisplay{\belowdisplayskip\new@belowdisplayshortskip\belowdisplayshortskip\new@belowdisplayshortskip}
\makeatother
\AtBeginDocument{\belowdisplayshortskip\belowdisplayskip}

\begin{document}

\maketitle

\begin{abstract}
For smooth convex disks $A$, i.e., convex compact subsets of the plane with non-empty interior and with at most one tangent at every boundary point, we classify the classes $G^{\text{hom}}(A)$ and $G^{\text{sim}}(A)$ of intersection graphs that can be obtained from homothets and similarities of $A$, respectively.
Namely, we prove that $G^{\text{hom}}(A)=G^{\text{hom}}(B)$ if and only if $A$ and $B$ are affine equivalent, and $G^{\text{sim}}(A)=G^{\text{sim}}(B)$ if and only if $A$ and $B$ are similar.
\end{abstract}

\section{Introduction}

Disk graphs have received much attention due to their ability to model graphs appearing in practice and their interesting structural properties.
In a disk graph, each vertex corresponds to a (circular) disk, and there is an edge between two vertices if and only if the two corresponding disks intersect.
Disk graphs appear naturally in problems related to radio and sensor networks.
For instance, the region reached by the signal from each transmitter in a radio network can be modeled as a disk, and when two disks intersect, the interference of the signals may be an issue if the transmitters use the same frequency.
The problem of avoiding interference while minimizing the number of used frequencies thus corresponds to finding the chromatic number of the disk graph.
Applications like that are part of the motivation for various papers on algorithms or computational hardness for problems taking disk graphs in the input \cite{alber2004geometric, baumann2024dynamic, bonamy2021eptas, caragiannis2007tight, clark1990unit, gibson2010algorithms, graf1998coloring, thai2007approximation} as well as papers studying disk graphs from a mathematical angle~\cite{mcdiarmid2013integer, mcdiarmid2014number}.

Combinatorial analysis of problems such as chromatic number and minimum hitting set size has often been performed in greater generality, for intersection graphs of translated copies or homothetic (i.e., translated and scaled) copies of a fixed convex shape \cite{dumitrescu2011piercing, kim2004chromatic, kim2006transversal, perepelitsa2003bounds}, and recently also for translated, scaled, and rotated squares~\cite{caoduro2024packing}.
Algorithmic considerations have also been generalized in a similar way---Bonnet, Grelier, and Miltzow~\cite{bonnet2020maximum} studied the maximum clique problem and extended classic algorithms for disk graphs and unit disk graphs to intersection graphs of homothetic or translated copies of a fixed convex set.

A well-established line of research in discrete and computational geometry has been aiming at understanding the relationships between classes of geometric intersection graphs such as whether two classes are equal or whether one class is a subclass of another \cite{cabello2017refining, cardinal2018intersection, chaplick2018grid, chmel2023string, janson1992thresholds, jelinek2019grounded, kratochvil1994intersection}.
In view of the above-mentioned research, it is natural to investigate the relationships between the classes of intersection graphs of translated copies, homothetic copies, and copies by similarity (translation, scaling, and rotation) of a fixed convex shape.

\begin{figure}[t]
\centering
\includegraphics[page=1]{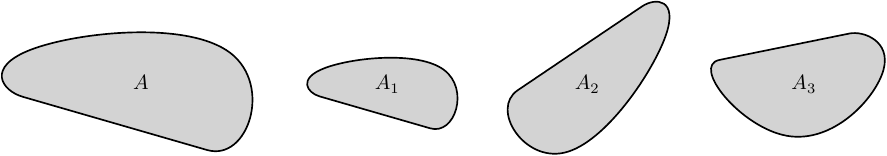}
\caption{Here, $A_1$ is a homothet of $A$, $A_2$ is a similarity but not a homothet of $A$, and $A_3$ is affine equivalent to $A$ but not similar to $A$.
By Theorem~\ref{thm:hom}, $A$ and $A_3$ induce the same intersection graphs of homothets, but Theorem~\ref{thm:sim} implies that the intersection graphs of similarities are different.}
\label{fig:disks}
\end{figure}

To be precise, consider an arbitrary \emph{convex disk} $A$, that is, a convex and compact set in the plane with non-empty interior.
A \emph{translate} of $A$ is a translated copy of $A$ (with no scaling or rotation allowed).
A \emph{homothet} of $A$ is a positively scaled and translated copy of $A$ (with no rotation allowed).
A \emph{similarity} is a homothet rotated by an arbitrary angle.
An \emph{affine equivalent} of $A$ is the image of $A$ under an invertible affine transformation.
See Figure~\ref{fig:disks}.
The \emph{intersection graph} of a family $\calF$ of sets in the plane is the graph with vertex set $\calF$ and edge set $\set{uv\colon u,v\in\calF,\:u\cap v\neq\emptyset}$.

In a recent paper, Aamand, Abrahamsen, Knudsen, and Rasmussen~\cite{aamand2021classifying} studied the question of when the translates of two convex disks induce the same intersection or contact graphs, where a \emph{contact graph} is an intersection graph that can be realized by disks with pairwise disjoint interiors.
They proved for a large class of convex disks, including all strictly convex ones, that two disks $A$ and $B$ yield the same classes of contact and intersection graphs if and only if the central symmetrals of $A$ and $B$ are affine equivalent, where the \emph{central symmetral} of a disk $A$ is the centrally symmetric disk $\frac{1}{2}(A-A)$.

In this paper, we study the question of when the homothets or the similarities of two convex disks induce the same intersection graphs.
We make the additional assumption that the convex disk $A$ be \emph{smooth}, that is, there is a unique tangent containing any point on the boundary of $A$.
We let $\hom A$ and $\sim A$ denote the sets of homothets and similarities of $A$, respectively.
We let $\Ghom(A)$ and $\Gsim(A)$ denote the classes of (finite) intersection graphs of homothets and similarities of $A$, respectively.
For two smooth convex disks $A$ and $B$, we are able to say exactly when $\Ghom(A)=\Ghom(B)$ and $\Gsim(A)=\Gsim(B)$, as follows.

\begin{theorem}\label{thm:hom}
Let\/ $A$ and\/ $B$ be smooth convex disks.
Then\/ $\Ghom(A)=\Ghom(B)$ if and only if\/ $A$ and\/ $B$ are affine equivalent.
Moreover, if\/ $A$ and\/ $B$ are not affine equivalent, then neither\/ $\Ghom(A)\subseteq\Ghom(B)$ nor\/ $\Ghom(B)\subseteq\Ghom(A)$.
\end{theorem}

\begin{theorem}\label{thm:sim}
Let\/ $A$ and\/ $B$ be smooth convex disks.
Then\/ $\Gsim(A)=\Gsim(B)$ if and only if\/ $B$~is~similar to\/ $A$ or to the reflection of\/ $A$.
\end{theorem}

If $A$ and $B$ are affine equivalent, then $\Ghom(A)=\Ghom(B)$, because the affine transformation that maps $A$ to $B$ transforms every realization in $\hom A$ to a realization of the same graph in $\hom B$, and vice versa.
Likewise, if $B$ is similar to $A$ or to the reflection of $A$, then $\Gsim(A)=\Gsim(B)$, because the similarity transformation (possibly with reflection) that maps $A$ to $B$ transforms every realization in $\sim A$ to a realization of the same graph in $\sim B$, and vice versa.
The difficult part is the necessity of these conditions.

When $A$ and $B$ are not affine equivalent, we exhibit graphs $G_A\in\Ghom(A)$ and $G_B\in\Ghom(B)$ such that $G_A\notin\Ghom(B)$ and $G_B\notin \Ghom(A)$, which yields the second part of Theorem~\ref{thm:hom}.
By contrast, when $B$ is dissimilar to both $A$ and the reflection of $A$, then $\Gsim(A)$ and $\Gsim(B)$ may be properly nested.
Indeed, if $A$ is a circular disk and $B$ is a non-circular filled ellipse, then $\Gsim(A)\subset\Gsim(B)$, because the affine stretch that maps $A$ to $B$ transforms every realization in $\hom A=\sim A$ to a realization of the same graph in $\hom B\subseteq\sim B$, while in the proof of Theorem~\ref{thm:sim}, we construct a graph in $\Gsim(B)$ that is not in $\Gsim(A)$.

One may or may not allow scaling by negative numbers when defining the homothets of $A$, which corresponds to rotating $A$ by $180$ degrees.
We remark that Theorem~\ref{thm:hom} holds in either case (with the same proof).
Likewise, one may or may not allow reflection when defining the similarities of $A$, and Theorem~\ref{thm:sim} holds in either case (with the same proof).

We note that although we establish results for more general families of graphs, our results are not generalizations of the ones in~\cite{aamand2021classifying}.
We also remark that the contact graphs of homothets or similarities of a smooth convex disk have already been characterized.
The Circle Packing Theorem, first proved by Koebe~\cite{koebe1936kontaktprobleme}, asserts that every planar graph is the contact graph of some set of pairwise interior-disjoint circular disks.
Since every contact graph is planar, the contact graphs are exactly the planar graphs.
Schramm's Packing Theorem \cite[Theorem~6.1]{schramm1996conformal} provides the following generalization.
Suppose that a planar graph is given, together with a correspondence which assigns a smooth convex disk to every vertex of the graph.
Then there exists a contact representation of the graph where every vertex is represented by a homothet of the associated disk.
As a consequence, the contact graphs of homothets or similarities of any smooth convex disk are the planar graphs.

\subsection*{Outline of the Paper}

In Sections \ref{sec:prelim} and~\ref{sec:convergence}, we set our notation and define basic concepts that we need further on.
In particular, in Section~\ref{sec:convergence}, we discuss Painlevé--Kuratowski convergence of subsets of $\setR^2$ and its properties.
It is more general than convergence based on the Hausdorff distance, as it allows us to express, for instance, that a sequence of (growing) convex disks converges to a half-plane.

\begin{figure}[t]
\centering
\includegraphics[page=2]{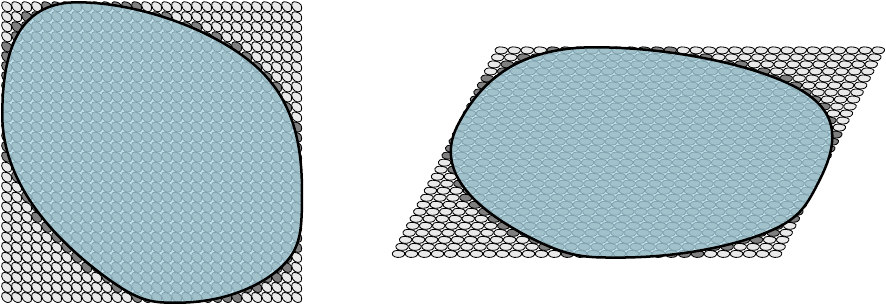}
\caption{To the left is shown the grid of small copies of $A$ and one large copy of $A$ on top.
The disks in the grid that are intersected (dark gray) define the shape of $A$ to an arbitrarily high precision, if we make the grid sufficiently fine.
To the right is shown the same graph realized by another disk $B$.
As we will show, the arrangement must again form a grid of small disks with one large copy of $B$ on top.
An affine map that makes the two grids coincide then also maps $B$ to $A$ to within a small error, since the two disks intersect the same ``pixels'' in the grids.}
\label{fig:grididea}
\end{figure}

In Sections \ref{sec:basic} and~\ref{sec:main}, we introduce constructions that enable us to distinguish the graph classes.
At an overall level, the idea behind our constructions is to define a graph $G$ such that however $G$ is realized as an intersection graph of homothets or similarities of a smooth convex disk $A$, a subset of the disks in the realization forms a large and almost regular grid of small copies of $A$; see Figure~\ref{fig:grididea}.
We use this grid in a somewhat similar manner as the grid of pixels in television.
We put one large disk $A$ on top of the grid.
The disks in the grid that intersect $A$ will then with high precision define the shape of $A$.
If now another disk $B$ is able to realize the same graph, then we can consider an affine transformation that makes the two grids ``match'', and it follows that $A$ and $B$ must be nearly identical under this transformation, since the same ``pixels'' in the two grids are intersected by the large disks on top.
If $B$ can realize the graph for an arbitrarily fine resolution of the grid, then we get in the limit a transformation $f^\star$ that maps $A$ to $B$.

In the case of homothets (Section~\ref{sec:homothets}), the transformation $f^\star$ is an arbitrary affine transformation, which leads to Theorem~\ref{thm:hom}.
In the case of similarities (Section~\ref{sec:similarities}), we can further prove that the grid must be square-shaped.
It then follows that the limit transformation $f^\star$ is angle preserving, so $B$ must be similar to $A$ or to the reflection of $A$.

The construction of this grid is rather delicate and relies on a careful analysis of various building blocks described in Section~\ref{sec:basic}.
Our first basic tool (Lemma~\ref{lemma:distance}) is that if the complete bipartite graph $K_{2,n}$ is realized as an intersection graph of similarities of a convex disk $A$, then the distance between the two disks $U_1$ and $U_2$ in the first vertex class can be made arbitrarily much smaller than the size of $U_1$ and $U_2$ by choosing $n$ large enough.
In other words, in the limit where $n\to\infty$, the two disks $U_1$ and $U_2$ behave as if they were in contact.

We are then able to define a larger graph $L_n$ where a realization has two disks $U_1,U_2$ and $n$ disks $V_1,\ldots,V_n$, such that by choosing $n$ large enough, we know that all the latter disks are arbitrarily small compared to both of $U_1$ and $U_2$ (Lemma~\ref{lemma:row}), and they must furthermore be ``squeezed in'' between these disks.
The disks in each row and each column of the aforementioned grid in the final construction will be a subset of the disks $V_1,\ldots,V_n$ in a realization of this graph $L_n$.
Here, it is necessary to place chains of overlapping disks on top of each row and each column of the grid to ensure that when the grid becomes arbitrarily fine, it does not degenerate into a segment.

In the case of similarities, we introduce the concept of \emph{stretch} for a convex disk $A$, denoted by $\rho_A$.
We consider two parallel lines at distance $1$ and a chain of $n$ consecutively overlapping similarities of $A$, contained in the strip bounded by these lines.
The stretch is the ratio between the (geometric) length of a longest such chain and $n$, as $n\to\infty$.
Now, if $\rho_B<\rho_A$, then it will be impossible for similarities of $B$ to realize the graph that we construct for $A$, as there is no chain of similarities of $B$ that can ``reach far enough''.
If $\rho_B=\rho_A$, then for both $A$ and $B$ the graph can be realized only so that the grid is square-shaped, since otherwise some chains in the realizations will not be able to reach far enough.

We conclude the paper in Section~\ref{sec:open} by mentioning some open questions.

\section{Preliminaries}\label{sec:prelim}

Let $\int X$ and $\partial X$ denote the interior and the boundary of a set $X\subseteq\setR^2$, respectively.
Two sets $X,Y\subseteq\setR^2$ are \emph{almost disjoint} if $X\cap\int Y=\emptyset$ and $\int X\cap Y=\emptyset$, and they \emph{touch} if they are almost disjoint but $X\cap Y\neq\emptyset$.
A \emph{convex disk} is a convex compact subset of $\setR^2$ with non-empty interior.
Every convex disk is the closure of its interior.
A \emph{tangent} to a convex disk $A$ is a straight line that touches $A$, whence it follows that $A$ lies in one of the two half-planes bounded by the line.
For every convex disk $A$ and every point $p\in\partial A$, there is at least one tangent to $A$ containing $p$.
A~convex disk $A$ is \emph{smooth} if for every $p\in\partial A$, there is exactly one tangent to $A$ containing $p$.

Let $\setR^+$ denote the set of positive reals.
Let $\origin$ denote the point $(0,0)\in\setR^2$.
Let $\rots$ denote the set of rotations in $\setR^2$ around $\origin$.
The set $\rots$ equipped with the metric induced by the operator norm forms a compact metric space such that the application $\rots\times\setR^2\ni(R,x)\mapsto R(x)\in\setR^2$ is continuous.
Let $\calH$ denote the family of half-planes in $\setR^2$.

A \emph{similarity} of a convex disk $A$ is a rotated, scaled, and translated copy of $A$, that is, a set of the form $A'=rR(A)+z=\set{rR(a)+z\colon a\in A}$ where $r\in\setR^+$, $R\in\rots$, and $z\in\setR^2$.
Let $r$ be called the \emph{radius} of $A'$ and denoted by $r_A(A')$ or simply by $r(A')$ when $A$ is clear from the context; it follows that $\area A'=r_A(A')^2\area A$.
A similarity $A'$ is a \emph{homothet} of $A$ if $R$ is the identity, that is, $A'$ is a scaled and translated copy of $A$.
Let $\sim A$ and $\hom A$ denote the set of similarities and the set of homothets of $A$, and let $\simrefl A=\sim A\cup\sim\refl{A}$, where $\refl{A}$ denotes the reflection of $A$ about the $y$ axis.

A \emph{realization} of a graph $G=(V,E)$ in a family $\calF$ of subsets of $\setR^2$ is a mapping $\rep\colon V\to\calF$ such that $\rep(u)\cap\rep(v)\neq\emptyset$ if and only if $uv\in E$.
We consider only finite graphs and their realizations with $\calF=\sim A$ or $\calF=\hom A$ for some convex disk $A$.

The Euclidean norm of a vector $a\in\setR^2$ is denoted by $\norm{a}$.
Thus, $\norm{p-q}$ is the Euclidean distance between points $p,q\in\setR^2$.
The distances between a point $p\in\setR^2$ and a non-empty set $X\subseteq\setR^2$ and between two non-empty sets $X,Y\subseteq\setR^2$ are defined as follows:
\[\dist(p,X)=\inf_{x\in X}\norm{p-x}{,}\qquad\dist(X,Y)=\inf_{x\in X}\inf_{y\in Y}\norm{x-y}{.}\]
The Hausdorff distance between non-empty subsets $X$ and $Y$ of $\setR^2$ is defined as follows:
\[d_H(X,Y)=\max\set[\bigg]{\sup_{x\in X}\dist(x,Y),\;\sup_{y\in Y}\dist(y,X)}{.}\]
For a positive real $\delta$, a point $x\in\setR^2$, and a compact set $X\subset\setR^2$, let
\[\disk{x}{\delta}=\set{p\in\setR^2\colon\norm{p-x}\leq\delta}{,}\qquad\disk{X}{\delta}=\set{p\in\setR^2\colon\dist(p,X)\leq\delta}{.}\]
The diameter of a set $X\subseteq\setR^2$, which is $\sup_{x,y\in X}\norm{x-y}$, is denoted by $\diam X$; it is a non-negative real number or infinity.
The \emph{bounding box} of a compact set $X\subset\setR^2$ is the unique minimal box of the form $[x_1,x_2]\times[y_1,y_2]$ containing $X$.

Let $\setN=\set{1,2,\ldots}$ and $[n]=\set{1,\ldots,n}$ for $n\in\setN$.
We allow sequences to be indexed by arbitrary infinite subsets of $\setN$, and we write the indices in the superscript.
For a sequence $(x^n)_{n\in N}$ with a suitable notion of convergence and a limit $x^\star$, we write $x^n\to x^\star$ \emph{on} $N$ to denote that $(x^n)_{n\in N}$ converges to $x^\star$.
A sequence $(X^n)_{n\in N}$ of non-empty subsets of $\setR^2$ is \emph{uniformly bounded} if $\sup_{n\in N}\sup_{x\in X^n}\norm{x}$ is finite, and it is \emph{anchored} if $\sup_{n\in N}\dist(\origin,X^n)$ is finite.

\section{Convergence and Limits}\label{sec:convergence}

Hausdorff distance is a metric on non-empty compact subsets of $\setR^2$ \cite[Kapitel~VIII, \S~6, II]{hausdorff1914grundzuge}, so it leads to a notion of convergence of sequences of such sets, in which the limit is also a non-empty compact subset of $\setR^2$.
A more general notion called Painlevé--Kuratowski convergence allows a sequence of compact sets to converge to an unbounded closed set such as a half-plane (whereas the Hausdorff distance between a bounded set and an unbounded set is always infinite).
For a sequence $(X^n)_{n\in N}$ of non-empty subsets of $\setR^2$, we define
\begin{align*}
\liminf\nolimits_{n\in N}X^n&=\set{x\in\setR^2\colon\dist(x,X^n)\to 0\text{ on }N}{,}\\
\limsup\nolimits_{n\in N}X^n&=\set{x\in\setR^2\colon\dist(x,X^n)\to 0\text{ on some infinite subset of }N}{.}
\end{align*}
Clearly, $\liminf_{n\in N}X^n\subseteq\limsup_{n\in N}X^n$, and both are closed sets.
We simply say that $(X^n)_{n\in N}$ \emph{converges} to a set $X^\star$, and we write $X^n\to X^\star$ \emph{on} $N$, if $\liminf_{n\in N}X^n=\limsup_{n\in N}X^n=X^\star$.
The following lemma relates the two notions of convergence on uniformly bounded sequences.

\begin{lemma}[{\cite[Kapitel~VIII, \S~6, IV and VI]{hausdorff1914grundzuge}}]
\label{lemma:convergence}
Let\/ $(X^n)_{n\in N}$ be a uniformly bounded sequence of non-empty subsets of\/ $\setR^2$, and let\/ $X^\star$ be a non-empty compact subset of\/ $\setR^2$.
Then\/ $X^n\to X^\star$ on\/ $N$ if and only if\/ $d_H(X^n,X^\star)\to 0$ on\/ $N$.
\end{lemma}

\begin{proof}
Suppose $d_H(X^n,X^\star)\to 0$ on $N$.
If $x\in X^\star$, then $\dist(x,X^n)\leq d_H(X^n,X^\star)\to 0$, so $x\in\liminf_{n\in N}X^n$, showing that $X^\star\subseteq\liminf_{n\in N}X^n$.
Now, let $x\in\limsup_{n\in N}X^n$, so that $\dist(x,X^n)\to 0$ on some infinite set $N'\subseteq N$.
For every $n\in N'$, there is $x^n\in X^n$ with $\norm{x-x^n}\leq\dist(x,X^n)+\frac{1}{n}$.
Consequently, we have $\dist(x,X^\star)\leq\norm{x-x^n}+\dist(x^n,X^\star)\leq\dist(x,X^n)+\frac{1}{n}+d_H(X^n,X^\star)\to 0$ on $N'$, which implies $x\in X^\star$ (as $X^\star$ is closed).
This shows that $\limsup_{n\in N}X^n\subseteq X^\star$.
Since $\liminf_{n\in N}X^n\subseteq\limsup_{n\in N}X^n$, we conclude that $\liminf_{n\in N}X^n=\limsup_{n\in N}X^n=X^\star$, that is, $X^n\to X^\star$ on $N$.

For the converse implication, suppose $X^n\to X^\star$ on $N$.
Suppose towards a contradiction that $\sup_{x\in X^n}\dist(x,X^\star)\not\to 0$ on $N$.
There are $\delta>0$ and an infinite set $N'\subseteq N$ such that for every $n\in N'$, we have $\sup_{x\in X^n}\dist(x,X^\star)>\delta$, so that there is $x^n\in X^n$ with $\dist(x^n,X^\star)\geq\delta$.
Since $(X^n)_{n\in N}$ is uniformly bounded, the sequence $(x^n)_{n\in N}$ is bounded.
Therefore, by the Bolzano--Weierstrass Theorem, we can restrict $N'$ further to an infinite subset on which $x^n\to x^\star\in\setR^2$.
This implies $x^\star\in\limsup_{n\in N}X^n=X^\star$ and consequently $\norm{x^n-x^\star}\geq\delta$ on $N'$, which is a contradiction.
Likewise, if $\sup_{y\in X^\star}\dist(y,X^n)\not\to 0$ on $N$, then there are $\delta>0$ and an infinite set $N'\subseteq N$ such that for every $n\in N'$, there is $y^n\in X^\star$ with $\dist(y^n,X^n)\geq\delta$.
Since $X^\star$ is compact, we have the following on some infinite subset of $N'$: $y^n\to y^\star\in X^\star=\liminf_{n\in N}X^n$ and thus $\dist(y^n,X^n)\leq\norm{y^n-y^\star}+\dist(y^\star,X^n)\to 0$, which is a contradiction.
We conclude that $d_H(X^n,X^\star)=\max\set{\sup_{x\in X^n}\dist(x,X^\star),\:\sup_{y\in X^\star}\dist(y,X^n)}\to 0$ on $N$.
\end{proof}

We remark that for all anchored sequences $(X^n)_{n\in N}$ of non-empty \emph{convex} compact subsets of $\setR^2$, Painlevé--Kuratowski convergence can be characterized as follows \cite[Theorem~4]{salinetti1979convergence}: $X^n\to X^\star$ on $N$ if and only if $d_H(X^n\cap\disk{\origin}{\rho},\:X^\star\cap\disk{\origin}{\rho})\to 0$ on $N$ for every $\rho\geq\sup_{n\in N}\dist(\origin,X^n)$.

\begin{lemma}
\label{lemma:two limits}
Let\/ $(X^n)_{n\in N}$ and\/ $(Y^n)_{n\in N}$ be sequences of non-empty subsets of\/ $\setR^2$.
\begin{enumerate}
\item\label{item:intersect} If\/ $(X^n)_{n\in N}$ is uniformly bounded and\/ $\dist(X^n,Y^n)\to 0$ on\/ $N$, then\/ $\limsup_{n\in N}X^n$ and\/ $\limsup_{n\in N}Y^n$ intersect.
\item\label{item:almost disjoint} If\/ $X^n$ and\/ $Y^n$ are disjoint and convex for all\/ $n\in N$, then\/ $\liminf_{n\in N}X^n$ and\/ $\liminf_{n\in N}Y^n$ are almost disjoint.
\end{enumerate}
\end{lemma}

\begin{proof}
For the proof of statement~\ref{item:intersect}, let $x^n\in X^n$ and $y^n\in Y^n$ be such that $\norm{x^n-y^n}\leq\dist(X^n,Y^n)+\frac{1}{n}$ for all $n\in N$.
Since $(X^n)_{n\in N}$ is uniformly bounded, the sequence $(x^n)_{n\in N}$ is bounded.
Therefore, by the Bolzano--Weierstrass Theorem, there is an infinite set $N'\subseteq N$ on which $x^n\to x^\star\in\setR^2$, which implies $x^\star\in\limsup_{n\in N}X^n$.
We also have
\[\norm{y^n-x^\star}\leq\norm{y^n-x^n}+\norm{x^n-x^\star}\leq\dist(X^n,Y^n)+\tfrac{1}{n}+\norm{x^n-x^\star}\to 0\text{ on }N'{,}\]
which implies $x^\star\in\limsup_{n\in N}Y^n$, showing that $(\limsup_{n\in N}X^n)\cap(\limsup_{n\in N}Y^n)\neq\emptyset$.

For the proof of statement~\ref{item:almost disjoint}, we need the following claim.

\begin{subclaim}
\label{claim:interior of liminf}
Let $(X^n)_{n\in N}$ be a sequence of non-empty convex subsets of $\setR^2$.
For every point $x\in\int(\liminf_{n\in N}X^n)$, there is $\epsilon>0$ such that $\disk{x}{\epsilon}\subseteq X^n$ for all but finitely many $n\in N$.
\end{subclaim}

\begin{claimproof}
Since $x\in\int(\liminf_{n\in N}X^n)$, there is $\epsilon>0$ with $\disk{x}{4\epsilon}\subseteq\liminf_{n\in N}X^n$.
Let $p_1$, $p_2$, $p_3$ be three equally spaced points on the circle with center $x$ and radius $4\epsilon$.
For each $i\in\set{1,2,3}$, since $p_i\in\liminf_{n\in N}X^n$, we have $\dist(p_i,X^n)<\epsilon$ for all but finitely many $n\in N$.
Therefore, for all but finitely many $n\in N$, there are points $x_1\in X^n\cap\disk{p_1}{\epsilon}$, $x_2\in X^n\cap\disk{p_2}{\epsilon}$, and $x_3\in X^n\cap\disk{p_3}{\epsilon}$, implying $\disk{x}{\epsilon}\subseteq X^n$ by convexity, as $\disk{x}{\epsilon}$ lies in the triangle $x_1x_2x_3$.
\end{claimproof}

Suppose $x\in\int(\liminf_{n\in N}X^n)\cap(\liminf_{n\in N}Y^n)$.
By Claim~\ref{claim:interior of liminf}, there is $\epsilon>0$ such that $\disk{x}{\epsilon}\subseteq X^n$ for all but finitely many $n\in N$.
Since $x\in\liminf_{n\in N}Y^n$, we have $\disk{x}{\epsilon}\cap Y^n\neq\emptyset$ for all but finitely many $n\in N$, which contradicts the assumption that $X^n\cap Y^n=\emptyset$ for all $n\in N$.
Thus $\int(\liminf_{n\in N}X^n)\cap(\liminf_{n\in N}Y^n)=\emptyset$, and analogously $(\liminf_{n\in N}X^n)\cap\int(\liminf_{n\in N}Y^n)=\emptyset$.
\end{proof}

A theorem of Zarankiewicz \cite[Théorème~1]{zarankiewicz1927points} asserts that every anchored sequence of non-empty subsets of $\setR^2$ contains a convergent subsequence.
We prove the following specialized variant of this statement.

\begin{lemma}
\label{lemma:limit}
Let\/ $A$ be a smooth convex disk.
Let\/ $\calF=\hom A$ or\/ $\calF=\sim A$.
Let\/ $(X^n)_{n\in N}$ be an anchored sequence of members of\/ $\calF$.
Then one of the following holds on some infinite subset of\/ $N$:
\begin{itemize}
\item $r(X^n)\to r^\star$ for some\/ $r^\star\in\setR^+$, and\/ $X^n\to X^\star$ for some\/ $X^\star\in\calF$ with\/ $r(X^\star)=r^\star$.
\item $r(X^n)\to 0$, and\/ $X^n\to\set{z^\star}$ for some point\/ $z^\star\in\setR^2$,
\item $r(X^n)\to\infty$, and\/ $X^n\to X^\star$ where\/ $X^\star$ is a half-plane or\/ $X^\star=\setR^2$.
\end{itemize}
Moreover, if\/ $(X^n)_{n\in N}$ is itself convergent, then one of the above holds on the entire set\/ $N$.
\end{lemma}

\begin{proof}
Let $(X^n)_{n\in N}$ be an anchored sequence of members of $\calF$ of the form $X^n=r^nR^n(A)+z^n$ where $r^n\in\setR^+$, $R^n\in\rots$ (the identity if $\calF=\hom A$), and $z^n\in\setR^2$ for all $n\in N$.

First, suppose that $r^n\to r^\star\in\setR^+\cup\set{0}$ on $N$.
Since the sequence $(X^n)_{n\in N}$ is anchored and $\diam(X^n\cup\set{z^n})=r^n\diam(A\cup\set{\origin})\to r^\star\diam(A\cup\set{\origin})$, the sequence $(z^n)_{n\in N}$ is bounded.
Therefore, by the Bolzano--Weierstrass Theorem and by compactness of $\rots$, there is an infinite set $N'\subseteq N$ on which $R^n\to R^\star\in\rots$ (the identity if $\calF=\hom A$) and $z^n\to z^\star\in\setR^2$.
Let $X^\star=r^\star R^\star(A)+z^\star$; it is a member of $\calF$ when $r^\star\in\setR^+$ or the singleton set $\set{z^\star}$ when $r^\star=0$.
Since the map $\setR^+\times\rots\times\setR^2\ni(r,R,z)\mapsto rR(a)+z\in\setR^2$ is continuous for every $a\in A$, it follows that $d_H(X^n,X^\star)\to 0$ and thus (by Lemma~\ref{lemma:convergence}) $X^n\to X^\star$ on $N'$.

Now, suppose that $r^n\to\infty$ on $N$.
If $\liminf_{n\in N}X^n=\setR^2$, then $X^n\to\setR^2$ on $N$, so suppose otherwise.
Let $p\in\setR^2\setminus\liminf_{n\in N}X^n$.
There are $\delta>0$ and an infinite set $N'\subseteq N$ with $\dist(p,X^n)\geq\delta$ for all $n\in N'$.
Let $x^n\in X^n$ be such that $\dist(p,X^n)=\norm{p-x^n}$ for $n\in N'$; such points exist as $X^n$ is compact.
Since the sequence $(X^n)_{n\in N}$ is anchored, the sequence $(x^n)_{n\in N'}$ is bounded.
Therefore, by the Bolzano--Weierstrass Theorem, we can restrict $N'$ further to an infinite subset on which $x^n\to x^\star\in\setR^2$, where $\norm{p-x^\star}\geq\delta$.
Let $H$ be the half-plane with $p\notin H$, $x^\star\in\partial H$, and $\partial H$ perpendicular to the segment $px^\star$, so that $\dist(p,H)=\norm{p-x^\star}$.

Let $A^n=(r^n)^{-1}(X^n-x^n)+x^\star=R^n(A)+c^n$ where $c^n=x^\star+(r^n)^{-1}(z^n-x^n)$ for $n\in N'$.
It~follows that $x^\star\in A^n$ for all $n\in N'$.
Since
\[\norm{c^n-x^\star}=(r^n)^{-1}\norm{z^n-x^n}\leq(r^n)^{-1}\diam(X^n\cup\set{z^n})=\diam(A\cup\set{\origin}){,}\]
the sequence $(c^n)_{n\in N'}$ is bounded.
Therefore, by the Bolzano--Weierstrass Theorem and by compactness of $\rots$, we can restrict $N'$ further to an infinite subset on which $R^n\to R^\star\in\rots$ and $c^n\to c^\star\in\setR^2$.
Let $A^\star=R^\star(A)+c^\star$.
Since the map $\rots\times\setR^2\ni(R,c)\mapsto R(a)+c\in\setR^2$ is continuous for every $a\in A$, it follows that $d_H(A^n,A^\star)\to 0$ and thus (by Lemma~\ref{lemma:convergence}) $A^n\to A^\star$ on $N'$.
Since $x^\star\in A^n$ for every $n\in N'$, it follows that $x^\star\in A^\star$.
We prove that $X^n\to H$ on $N'$.

\begin{subclaim}
\label{claim:halfplane-1}
For every $x\in\setR^2$, if the segment $xx^\star$ intersects $A^\star$ at some point other than $x^\star$, then $x\in\liminf_{n\in N'}X^n$.
\end{subclaim}

\begin{claimproof}
Suppose the segment $xx^\star$ intersects $A^\star$ at some point $a^\star\neq x^\star$.
Let $r=\norm{x-x^\star}/\norm{a^\star-x^\star}$, so that $x=x^\star+r(a^\star-x^\star)$.
Since $A^n\to A^\star$ on $N'$, there is a sequence $(a^n)_{n\in N'}$ with $A^n\ni a^n\to a^\star$ on $N'$.
For each $n\in N'$, let $\hat x^n\in X^n$ satisfy $a^n=x^\star+(r^n)^{-1}(\hat x^n-x^n)$, and let $y^n=x^n+r(r^n)^{-1}(\hat x^n-x^n)$, which belongs to $X^n$ by convexity when $r^n\geq r$.
It follows that
\begin{align*}
\norm{y^n-x}&=\norm{(x^n+r(r^n)^{-1}(\hat x^n-x^n))-(x^\star+r(a^\star-x^\star))}=\norm{(x^n-x^\star)+r(a^n-a^\star)}\\
&\leq\norm{x^n-x^\star}+r\cdot\norm{a^n-a^\star}\to 0\text{ on }N'{,}
\end{align*}
as $x^n\to x^\star$ and $r^n\to\infty$ on $N'$.
We conclude that $x\in\liminf_{n\in N'}X^n$.
\end{claimproof}

\begin{subclaim}
\label{claim:halfplane-2}
$\limsup_{n\in N'}X^n\subseteq H$.
\end{subclaim}

\begin{claimproof}
Suppose $y^\star\in(\limsup_{n\in N'}X^n)\setminus H$.
It follows that there are an infinite set $N''\subseteq N'$ and a sequence $(y^n)_{n\in N''}$ with $X^n\ni y^n\to y^\star$ on $N''$.
Since $\partial H$ is perpendicular to the segment $px^\star$ and $y^\star\notin H$, the segment $x^\star y^\star$ contains a point $z^\star=\lambda x^\star+(1-\lambda)y^\star$ with $\lambda\in[0,1]$ and $\norm{p-z^\star}<\norm{p-x^\star}$.
Let $\epsilon=\norm{p-x^\star}-\norm{p-z^\star}>0$.
Since $x^n\to x^\star$ and $y^n\to y^\star$ on $N''$, there is $n\in N''$ such that $\norm{x^n-x^\star}<\frac{\epsilon}{2}$ and $\norm{y^n-y^\star}<\frac{\epsilon}{2}$.
Let $z^n=\lambda x^n+(1-\lambda)y^n$.
We thus have
\begin{gather*}
\norm{z^n-z^\star}=\norm{\lambda(x^n-x^\star)+(1-\lambda)(y^n-y^\star)}\leq\lambda\norm{x^n-x^\star}+(1-\lambda)\norm{y^n-y^\star}<\tfrac{\epsilon}{2}{,}\\
\begin{aligned}
\norm{p-z^n}&\leq\norm{p-z^\star}+\norm{z^\star-z^n}<\norm{p-z^\star}+\tfrac{\epsilon}{2}=\norm{p-x^\star}-\tfrac{\epsilon}{2}\\
&<\norm{p-x^\star}-\norm{x^\star-x^n}\leq\norm{p-x^n}=\dist(p,X^n){.}
\end{aligned}
\end{gather*}
However, $z^n\in X^n$ by convexity, which is a contradiction.
\end{claimproof}

By Claims \ref{claim:halfplane-1} and~\ref{claim:halfplane-2}, we have $A^\star\subseteq\liminf_{n\in N'}X^n\subseteq\limsup_{n\in N'}X^n\subseteq H$.
Since $x^\star\in A^\star\cap\partial H$, it follows that $\partial H$ is a tangent to $A^\star$ at $x^\star$.
Moreover, it is a unique such tangent, as $A^\star$ is a smooth convex disk.
Consequently, for every $x\in\int H$, the segment $xx^\star$ intersects $\int A^\star$, which implies that $\int H\subseteq\liminf_{n\in N'}X^n$, by Claim~\ref{claim:halfplane-1}.
Since $\liminf_{n\in N'}X^n$ is closed, we conclude that $H\subseteq\liminf_{n\in N'}X^n\subseteq\limsup_{n\in N'}X^n\subseteq H$, so $X^n\to H$ on $N'$.

Consider now the general setting of the lemma.
There is an infinite subset of $N$ on which $r^n\to r^\star\in\setR^+\cup\set{0}$ or $r^n\to\infty$.
Either case leads to an infinite subset of $N$ on which $X^n$ converges to a limit claimed by the lemma, as we have shown above.
Moreover, if neither $r^n\to r^\star\in\setR^+\cup\set{0}$ nor $r^n\to\infty$ on $N$, then there are infinite subsets of $N$ on which $r^n$ tends to different limits, leading to infinite subsets of $N$ with different limits of $X^n$.
The last statement of the lemma now follows by the fact that if $X^n\to X^\star$ on $N$, then $X^n\to X^\star$ on every infinite subset of $N$.
\end{proof}

\begin{lemma}
\label{lemma:sequence from limit}
Let\/ $A$ be a smooth convex disk, and let\/ $\calF=\hom A$ or\/ $\calF=\sim A$.
For every set\/ $X^\star\in\calF\cup\calH$, there is a sequence\/ $(X^n)_{n\in\setN}$ of members of\/ $\calF$ such that\/ $X^n\subset\int X^\star$ for all\/ $n\in\setN$ and\/ $X^n\to X^\star$.
\end{lemma}

\begin{proof}
If $X^\star\in\calF$, then taking $X^n=(1-\frac{1}{n+1})(X^\star-q)+q$ for an arbitrary point $q\in\int X^\star$ and all $n\in\setN$ yields a sequence $(X^n)_{n\in\setN}$ of homothets of $X^\star$ with $X^n\subset\int X^\star$ and $d_H(X^n,X^\star)\to 0$, the latter implying $X^n\to X^\star$ by Lemma~\ref{lemma:convergence}.

Now, suppose $X^\star$ is a half-plane.
Let $X$ be a translate of $A$ contained in $X^\star$ and tangent to $\partial X^\star$ at a point $p$.
Let $q\in\int X$.
Let $X^n=n(X-p)+p+\frac{1}{n}(q-p)$ for $n\in\setN$.
Thus $X^n\subset\int X^\star$ for all $n\in\setN$, which implies $\limsup_{n\in\setN}X^n\subseteq X^\star$.
Since $A$ is smooth, $\partial X^\star$ is the unique tangent to $X$ at $p$.
Therefore, for every $x\in\int X^\star$, the intersection of $X$ with the segment $px$ has positive length, which implies that $x\in n(X-p)+p$ and thus $x+\frac{1}{n}(q-p)\in X^n$ for all but finitely many $n\in\setN$, showing that $\dist(x,X^n)\to 0$.
Consequently, $\int X^\star\subseteq\liminf_{n\in\setN}X^n$.
Since the latter set is closed, we conclude that $X^\star\subseteq\liminf_{n\in\setN}X^n\subseteq\limsup_{n\in\setN}X^n\subseteq X^\star$, so $X^n\to X^\star$.
\end{proof}

An \emph{interior-realization} of a graph $G=(V,E)$ in a family $\bar\calF$ of subsets of $\setR^2$ is a mapping $\intrep\colon V\to\bar\calF$ such that $\int\intrep(u)\cap\int\intrep(v)\neq\emptyset$ if and only if $uv\in E$.
Our main construction in Section~\ref{sec:main} is easier to present in terms of interior-realizations rather than realizations, and the following lemma turns an interior-realization into a realization.

\begin{lemma}
\label{lemma:interior-realization}
Let\/ $A$ be a smooth convex disk and\/ $\calF=\hom A$ or\/ $\calF=\sim A$.
If a graph\/ $G$ has an interior-realization in\/ $\calF\cup\calH$, then\/ $G$ has a realization in\/ $\calF$.
\end{lemma}

\begin{proof}
Let $G=(V,E)$, and let $\intrep$ be an interior-realization of $G$ in $\calF\cup\calH$.
Let $p_{uv}\in\int\intrep(u)\cap\int\intrep(v)$ for every edge $uv\in E$.
By Lemma~\ref{lemma:sequence from limit}, there are mappings $\rep^n\colon V\to\calF$ for $n\in\setN$ such that $\int\intrep(v)\supset\rep^n(v)\to\intrep(v)$ for every $v\in V$.
It follows that $\rep^n(u)\cap\rep^n(v)\neq\emptyset$ implies $\int\intrep(u)\cap\int\intrep(v)\neq\emptyset$ and thus $uv\in E$, for all $n\in\setN$.
If $n\in\setN$ is sufficiently large that $p_{uv}\in\rep^n(u)\cap\rep^n(v)$ for every edge $uv\in E$, then $\rep^n$ is a realization of $G$ in $\calF$.
\end{proof}

\section{Basic Configurations}\label{sec:basic}

\begin{lemma}
\label{lemma:star}
For every convex disk\/ $A$ and every\/ $\epsilon>0$, if\/ $n\in\setN$ is sufficiently large, then for any disks\/ $V_1,\ldots,V_n\in\sim A$ that are pairwise almost disjoint and all intersect a common disk\/ $U\in\sim A$, we have\/ $\min_{i\in[n]}r(V_i)<\epsilon r(U)$.
\end{lemma}

\begin{proof}
Let $n>\alpha/(\epsilon^2\area A)$ where $\alpha=\area\disk{A}{\epsilon\diam A}$.
Let $\rho=\epsilon r(U)$, and suppose towards a contradiction that $\min_{i\in[n]}r(V_i)\geq\rho$.
Let $D=\disk{U}{\rho\diam A}$.
We claim that $\area(D\cap V_i)\geq\rho^2\area A$ for every $i\in[n]$.
Indeed, if $V_i\subseteq D$, then $\area(D\cap V_i)=\area V_i=r(V_i)^2\area A\geq\rho^2\area A$, and otherwise the set $D\cap V_i$ contains a homothet of $V_i$ (so a member of $\sim A$) intersecting $U$ and touching $\partial D$, which therefore has diameter at least $\rho\diam A$ and area at least $\rho^2\area A$.
Since $V_1,\ldots,V_n$ are pairwise almost disjoint, it follows that
\[\area D=r(U)^2\alpha<n\rho^2\area A\leq\sum_{i\in[n]}\area(D\cap V_i)=\area\biggl(D\cap\bigcup_{i\in[n]}V_i\biggr)\leq\area D{.}\]
This contradiction shows that $\min_{i\in[n]}r(V_i)<\rho=\epsilon r(U)$.
\end{proof}

Let $K_{2,n}$ denote the complete bipartite graph with vertices $u_1$ and $u_2$ on one side and $v_1,\ldots,v_n$ on the other side, so that $u_iv_j$ is an edge of $K_{2,n}$ for all $i\in[2]$ and $j\in[n]$.
We say that the vertices $u_1$ and $u_2$ form the \emph{small side} of the graph $K_{2,n}$.

\begin{lemma}
\label{lemma:distance}
Let\/ $A$ be a convex disk.
Let\/ $(\rep^n)_{n\in N}$ be a sequence such that\/ $\rep^n$ is a realization of\/ $K_{2,n}$ in\/ $\sim A$ and\/ $(\rep^n(u_1))_{n\in N}$ converges to a convex disk or a singleton set\/ $U_1^\star$.
Then\/ $\dist(\rep^n(u_2),U_1^\star)\to 0$ on\/ $N$, so in particular the sequence\/ $(\rep^n(u_2))_{n\in N}$ is anchored.
Moreover, let\/ $P$ be the set of points\/ $p\in\setR^2$ for which there are an infinite set\/ $N'\subseteq N$ and a sequence\/ $(W^n)_{n\in N'}$ with\/ $W^n\in\set{\rep^n(v_1),\ldots,\rep^n(v_n)}$ and\/ $W^n\to\set{p}$ on\/ $N'$.
Then\/ $P$ is non-empty, and if\/ $(\rep^n(u_2))_{n\in N}$ converges to a set\/ $U_2^\star$, then\/ $U_1^\star$ and\/ $U_2^\star$ touch at every point\/ $p\in P$.
\end{lemma}

\begin{proof}
For every $n\in N$, let $V^n\in\set{\rep^n(v_1),\ldots,\rep^n(v_n)}$ be such that $r(V^n)=\min_{i\in[n]}r(\rep^n(v_i))$.
By Lemma~\ref{lemma:limit}, $r(\rep^n(u_1))\to r^\star\in\setR^+\cup\set{0}$, which implies $r(V^n)\to 0$ on $N$, by Lemma~\ref{lemma:star}.
Thus
\[\dist(\rep^n(u_1),\rep^n(u_2))\leq\diam V^n=r(V^n)\cdot\diam A\to 0\text{ on }N{.}\]
Since $\rep^n(u_1)\to U_1^\star$ on $N$, Lemma~\ref{lemma:convergence} yields $d_H(\rep^n(u_1),U_1^\star)\to 0$ on $N$.
Consequently,
\[\dist(\rep^n(u_2),U_1^\star)\leq\dist(\rep^n(u_2),\rep^n(u_1))+\sup_{x\in\rep^n(u_1)}\dist(x,U_1^\star)\to 0\text{ on }N{.}\]
Since $r(V^n)\to 0$ on $N$, by Lemma~\ref{lemma:limit}, there is an infinite set $N'\subseteq N$ on which $V^n\to\set{p}$ for some point $p$ which (by definition) belongs to $P$.
In particular, $P$ is non-empty.

Now, let $p\in P$, so that there are an infinite set $N'\subseteq N$ and a sequence $(W^n)_{n\in N'}$ with $W^n\in\set{\rep^n(v_1),\ldots,\rep^n(v_n)}$ and $W^n\to\set{p}$ on $N'$.
Suppose that $(\rep^n(u_2))_{n\in N}$ converges to some set $U_2^\star$.
Let $w_i^n\in W^n\cap\rep^n(u_i)$ for $i\in[2]$ and $n\in N'$.
It follows that $w_i^n\to p$ on $N'$, which implies $p\in\limsup_{n\in N}\rep^n(u_i)=U_i^\star$ for $i\in[2]$.
By Lemma~\ref{lemma:two limits}~(\ref{item:almost disjoint}), the sets $U_1^\star$ and $U_2^\star$ are almost disjoint, so they touch at $p$.
\end{proof}

\begin{figure}[t]
\centering
\includegraphics[page=3]{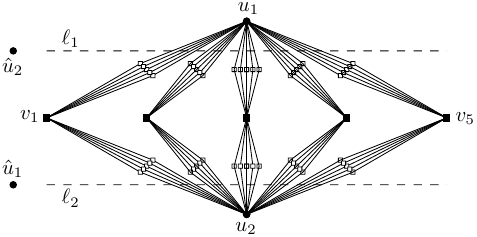}
\caption{The graph $L_5$.
Here, $\hat u_1$ has an edge to all vertices above the line $\ell_2$, and $\hat u_2$ has an edge to all vertices below $\ell_1$.}
\label{fig:L5}
\end{figure}

\begin{construction}[the graph $L_n$]
The graph $L_n$ has vertices $u_1,u_2,v_1,\ldots,v_n$, vertices $w_{ijk}$ and edges $u_iw_{ijk},w_{ijk}v_j$ for all $i\in[2]$ and $j,k\in[n]$ (so that $u_i,v_j,w_{ij1},\ldots,w_{ijn}$ form a copy of $K_{2,n}$), and two additional vertices $\hat u_1,\hat u_2$ such that $\hat u_1$ has an edge to every vertex except $u_2$ and $\hat u_2$ has an edge to every vertex except $u_1$.
See Figure~\ref{fig:L5}.
\end{construction}

When considering a specific realization $\rep$ of $L_n$ (possibly with a superscript), we write $V_i$, $U_i$, and $\hat U_i$ (with the same superscript) as shorthand for $\rep(v_i)$, $\rep(u_i)$, and $\rep(\hat u_i)$, respectively.

\begin{lemma}
\label{lemma:row}
For every smooth convex disk\/ $A$ and every\/ $\epsilon>0$, if\/ $n$ is sufficiently large, then every realization of\/ $L_n$ in\/ $\sim A$ satisfies\/ $\max_{j\in[n]}r(V_j)\leq\epsilon\min\set{r(U_1),r(U_2)}$.
\end{lemma}

\begin{proof}
Suppose towards a contradiction that there are $\epsilon>0$ and an infinite set $N\subseteq\setN$ such that for every $n\in N$, there is a realization $\rep^n$ of $L_n$ in $\sim A$ with $\max_{j\in[n]}r(V_j^n)>\epsilon\min\set{r(U_1^n),r(U_2^n)}$.
Assume without loss of generality that $r(U_1^n)\leq r(U_2^n)$ for all $n\in N$.
Furthermore, assume that $U_1^n$ is constant (equal to $U_1$) while the other disks may change size and placement as a function of $n$.

Suppose towards a contradiction that there is $\rho>0$ with $\min_{i\in[n]}r(V_i^n)\geq\rho$ for infinitely many $n\in N$.
Let $k\in N$.
By Lemmas \ref{lemma:limit} and~\ref{lemma:distance}, we can restrict $N$ further to an infinite subset on which $n\geq k$ and $V_i^n$ converges to a limit $V_i^\star\in\sim A\cup\calH$ that touches $U_1$ for every $i\in[k]$; moreover, $r(V_i^\star)\geq\rho$ whenever $V_i^\star\in\sim A$.
By Lemma~\ref{lemma:two limits}~(\ref{item:almost disjoint}), since $V_1^n,\ldots,V_k^n$ are pairwise disjoint for every $n\geq k$, the limits $V_1^\star,\ldots,V_k^\star$ are pairwise almost disjoint.
At most two of these limits can be half-planes, so at least $k-2$ of them are members of $\sim A$.
Moreover, since they all intersect $U_1$, by Lemma~\ref{lemma:star}, at least one of them has radius less than $\rho$ for sufficiently large $k$.
This contradiction shows that $\min_{i\in[n]}r(V_i^n)\to 0$ on $N$.

For each $n\in N$, let $V_{\min}^n$ and $V_{\max}^n$ be disks among $V_1^n,\ldots,V_n^n$ with minimum and maximum radii, respectively, so that $r(V_{\max}^n)>\epsilon r(U_1)$ and $r(V_{\min}^n)\to 0$ on $N$.
See Figure~\ref{fig:lemma:row}.
By Lemmas \ref{lemma:limit} and~\ref{lemma:distance}, we can restrict $N$ further to an infinite subset on which
\begin{itemize}
\item $U_2^n$ converges to a limit $U_2^\star\in\sim A\cup\calH$, as $r(U_2^n)\geq r(U_1)$ for all $n\in N$;
\item $V_{\min}^n$ converges to a singleton set $\set{p}$ that touches $U_1$ and $U_2^\star$, so that $p\in U_1\cap U_2^\star$;
\item $\hat U_1^n$ and $\hat U_2^n$ converge to limits $\hat U_1^\star$ and $\hat U_2^\star$, respectively, with $p\in\hat U_1^\star\cap\hat U_2^\star$, by Lemma~\ref{lemma:two limits}~(\ref{item:intersect});
\item $V_{\max}^n$ converges to a limit $V_{\max}^\star\in\sim A\cup\calH$ that touches $U_1$ and $U_2^\star$.
\end{itemize}
By Lemma~\ref{lemma:two limits}~(\ref{item:almost disjoint}), the three pairs of sets $U_1$ and $U_2^\star$, $\hat U_1^\star$ and $U_2^\star$, and $U_1$ and $\hat U_2^\star$ are almost disjoint.
Since $p\in U_1\cap U_2^\star\cap\hat U_1^\star\cap\hat U_2^\star$, it follows that the two sets in each of the three pairs touch at $p$.
Since $A$ is smooth, the unique straight line tangent to both $U_1$ and $U_2^\star$ at $p$ splits the plane into two half-planes $H_1$ and $H_2$ such that $U_1,\hat U_1^\star\subseteq H_1$ and $U_2^\star,\hat U_2^\star\subseteq H_2$.

\begin{figure}[t]
\centering
\includegraphics[page=4]{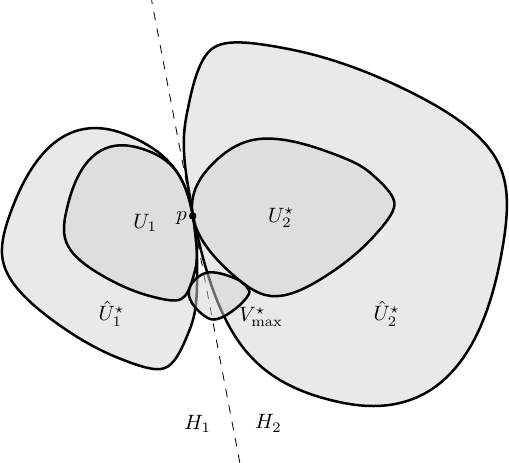}
\caption{Situation from the proof of Lemma~\ref{lemma:row}.}
\label{fig:lemma:row}
\end{figure}

Suppose that at least one of $U_2^\star$, $V_{\max}^\star$ is a member of $\sim A$.
By Lemma~\ref{lemma:star}, for $n\in N$, there are disks $W_1^n$ and $W_2^n$ among $\rep^n(w_{1jk})$ and $\rep^n(w_{2jk})$, respectively, such that
\begin{itemize}
\item $W_1^n$ intersects $V_{\max}^n$, $U_1$, and $\hat U_2^n$,
\item $W_2^n$ intersects $V_{\max}^n$, $U_2^n$, and $\hat U_1^n$,
\item $r(W_1^n)\to 0$ and $r(W_2^n)\to 0$ on $N$.
\end{itemize}
Since the sequences $(W_1^n)_{n\in N}$ and $(W_2^n)_{n\in N}$ are anchored, by Lemma~\ref{lemma:limit}, we can restrict $N$ further to an infinite subset on which $W_1^n\to\set{q_1}$ and $W_2^n\to\set{q_2}$ where $q_1\in V_{\max}^\star\cap U_1\cap\hat U_2^\star$ and $q_2\in V_{\max}^\star\cap\hat U_1^\star\cap U_2^\star$.
It follows that $V_{\max}^\star$ touches $U_1$ at $q_1$ and $U_2^\star$ at $q_2$, whereas both $q_1$ and $q_2$ lie on the boundary line between $H_1$ and $H_2$.
This is possible only when $V_{\max}^\star=\set{q_1}=\set{q_2}$, which is a contradiction.

Now, suppose that both $U_2^\star$ and $V_{\max}^\star$ are half-planes (so $U_2^\star=H_2$).
By Lemma~\ref{lemma:two limits}~(\ref{item:almost disjoint}), $U_1$, $U_2^\star$, and $V_{\max}^\star$ are pairwise almost disjoint, which implies that $U_2^\star$ and $V_{\max}^\star$ are disjoint half-planes.
However, $\hat U_2^\star\subseteq H_2=U_2^\star$, so $V_{\max}^\star$ and $\hat U_2^\star$ are disjoint, which is again a contradiction.
\end{proof}

\begin{lemma}
\label{lemma:extended row}
Let\/ $A$ be a smooth convex disk, let\/ $m\in\setN$, and let\/ $N$ be an infinite subset of\/ $\setN$ with all members at least\/ $m$.
For each\/ $n\in N$, let\/ $L'_n$ be a graph which contains, as induced subgraphs, $L_n$ and a fixed connected graph\/ $H$ containing\/ $v_1,\ldots,v_m$ such that\/ $u_1$ and\/ $u_2$ have no edges to any vertex of\/ $H$.
Let\/ $(\rep^n)_{n\in N}$ be a sequence such that\/ $\rep^n$ is a realization of\/ $L'_n$ in\/ $\sim A$ and\/ $V_1^n$ converges to a convex disk\/ $V_1^\star$ on\/ $N$.
Then\/ $N$ has an infinite subset on which
\begin{itemize}
\item $U_1^n$ and\/ $U_2^n$ converge to disjoint half-planes\/ $U_1^\star$ and\/ $U_2^\star$,
\item $\hat U_1^n$ and\/ $\hat U_2^n$ converge to limits that touch\/ $U_2^\star$ and\/ $U_1^\star$, respectively,
\item for every vertex\/ $w$ of\/ $H$, $\rep^n(w)$ converges to a convex disk or a singleton set,
\item $V_1^n,\ldots,V_m^n$ converge to convex disks\/ $V_1^\star,\ldots,V_m^\star$ that all touch\/ $U_1^\star$ and\/ $U_2^\star$.
\end{itemize}
\end{lemma}

\begin{proof}
By Lemma~\ref{lemma:distance}, the sequences $(U_1^n)_{n\in N}$ and $(U_2^n)_{n\in N}$ are anchored, and so are the sequences $(\hat U_1^n)_{n\in N}$ and $(\hat U_2^n)_{n\in N}$.
Therefore, by Lemma~\ref{lemma:limit}, we can restrict $N$ to an infinite subset on which $U_1^n$, $U_2^n$, $\hat U_1^n$, and $\hat U_2^n$ converge to limits $U_1^\star$, $U_2^\star$, $\hat U_1^\star$, and $\hat U_2^\star$, respectively.
By Lemma~\ref{lemma:distance}, $\hat U_1^\star$ touches $U_2^\star$, $\hat U_2^\star$ touches $U_1^\star$, and $V_1^\star$ touches both $U_1^\star$ and $U_2^\star$.
Since $V_1^\star$ is a convex disk, Lemma~\ref{lemma:limit} implies that $r(V_1^n)\to r(V_1^\star)$ on $N$, whence it follows by Lemma~\ref{lemma:row} that $r(U_1^n)\to\infty$ and $r(U_2^n)\to\infty$.
Therefore, by Lemma~\ref{lemma:limit}, $U_1^\star$ and $U_2^\star$ are half-planes.
The two half-planes are disjoint because $U_1^\star$, $U_2^\star$, and $V_1^\star$ are pairwise almost disjoint, by Lemma~\ref{lemma:two limits}~(\ref{item:almost disjoint}).

We claim that for every subset $X$ of the vertices of $H$ containing $v_1$ and inducing a connected subgraph of $H$, we can restrict $N$ further to an infinite subset on which $\rep^n(w)$ converges to a convex disk or a singleton set for every $w\in X$.
We prove the claim by induction on $\abs{X}$, noting that it is trivial when $\abs{X}=1$, that is, $X=\set{v_1}$.
Suppose that $\abs{X}\geq 2$ and the claim holds for all subsets of size $\abs{X}-1$.
There is a vertex $v\in X\setminus\set{v_1}$ such that the subgraph of $H$ induced on $X\setminus\set{v}$ is connected.
The set $X\setminus\set{v}$ contains a neighbor $w$ of $v$ in $H$.
By the induction hypothesis, we can assume that $\rep^n(w)$ converges to a convex disk or a singleton set $W^\star$.
This implies that the considered subsequence of $\rep^n(v)$ is anchored and therefore, by Lemma~\ref{lemma:limit}, it converges to a limit $V^\star$ after a suitable further restriction of $N$.
Since $\rep^n(v)$ is disjoint from $U_1^n$ and $U_2^n$ for all $n\in N$, by Lemma~\ref{lemma:two limits}~(\ref{item:almost disjoint}), $V^\star$ is disjoint from $\int U_1^\star$ and $\int U_2^\star$, so by Lemma~\ref{lemma:limit}, it is a convex disk or a singleton set.
This completes the induction step in the proof of the claim.

The claim applied to the whole vertex set of $H$ implies that $V_1^n,\ldots,V_m^n$ converge to convex disks or singleton sets $V_1^\star,\ldots,V_m^\star$.
By Lemma~\ref{lemma:distance}, the latter limits all touch $U_1^\star$ and $U_2^\star$.
Since the two half-planes are disjoint, $V_1^\star,\ldots,V_m^\star$ are not singleton sets---they are convex disks.
\end{proof}

\section{Main Construction}\label{sec:main}

\begin{figure}[t]
\centering
\includegraphics[page=5]{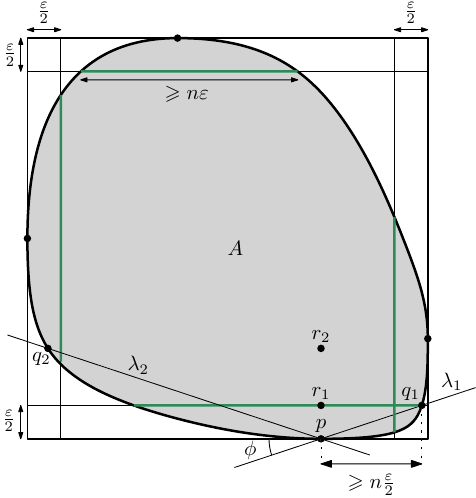}
\caption{Lemma~\ref{lemma:smoothness} asserts that for every $n\in\setN$, if $\epsilon>0$ is sufficiently small, then the lengths of the four green segments are at least $n\epsilon$.
Here, $\epsilon$ is realized as $2\cdot\norm{r_1-p}$.}
\label{fig:smoothness}
\end{figure}

The following simple lemma is illustrated in Figure~\ref{fig:smoothness}.

\begin{lemma}
\label{lemma:smoothness}
Let\/ $A$ be a smooth convex disk with bounding box\/ $[x_1,x_2]\times[y_1,y_2]$.
For every\/ $n\in\setN$, there is\/ $\epsilon_0>0$ such that for every\/ $\epsilon\in(0,\epsilon_0]$, each of the four segments\/ $A\cap(\setR\times\set{y_1+\frac{\epsilon}{2}})$, $A\cap(\setR\times\set{y_2-\frac{\epsilon}{2}})$, $A\cap(\set{x_1+\frac{\epsilon}{2}}\times\setR)$, and\/ $A\cap(\set{x_2-\frac{\epsilon}{2}}\times\setR)$ has length at least\/ $n\epsilon$.
\end{lemma}

\begin{proof}
Let $p\in A\cap(\setR\times\set{y_1})$, and let $\lambda_1$ and $\lambda_2$ be the lines through $p$ that make angles of $\phi=\arctan\frac{1}{n}$ with the horizontal line $\setR\times\set{y_1}$.
For each $i\in\set{1,2}$, there is a point $q_i\neq p$ such that $q_i\in\partial A\cap\lambda_i$, otherwise $\lambda_i$ and $\setR\times\set{y_1}$ would both be tangents to $A$ at $p$, contradicting smoothness of $A$.
Let $r_i$ be the intersection point of the vertical line through $p$ and the horizontal line through $q_i$.
We then have $\norm{q_i-r_i}/\norm{r_i-p}=1/\tan\phi=n$.
It follows that when $\epsilon\leq 2\min\set{\norm{r_1-p},\norm{r_2-p}}$, then the segment $A\cap(\setR\times\set{y_1+\frac{\epsilon}{2}})$ has length at least $n\epsilon$.

Similarly, we consider intersection points with the other sides of the bounding box $[x_1,x_2]\times[y_1,y_2]$, i.e., points in $A\cap(\setR\times\set{y_2})$, $A\cap(\set{x_1}\times\setR)$, and $A\cap(\set{x_2}\times\setR)$, and the lines that make angles of $\phi$ with these sides.
We then get further upper bounds on $\epsilon$, and let $\epsilon_0$ be the minimum of the four bounds.
\end{proof}

\begin{figure}[t]
\centering
\includegraphics[page=6]{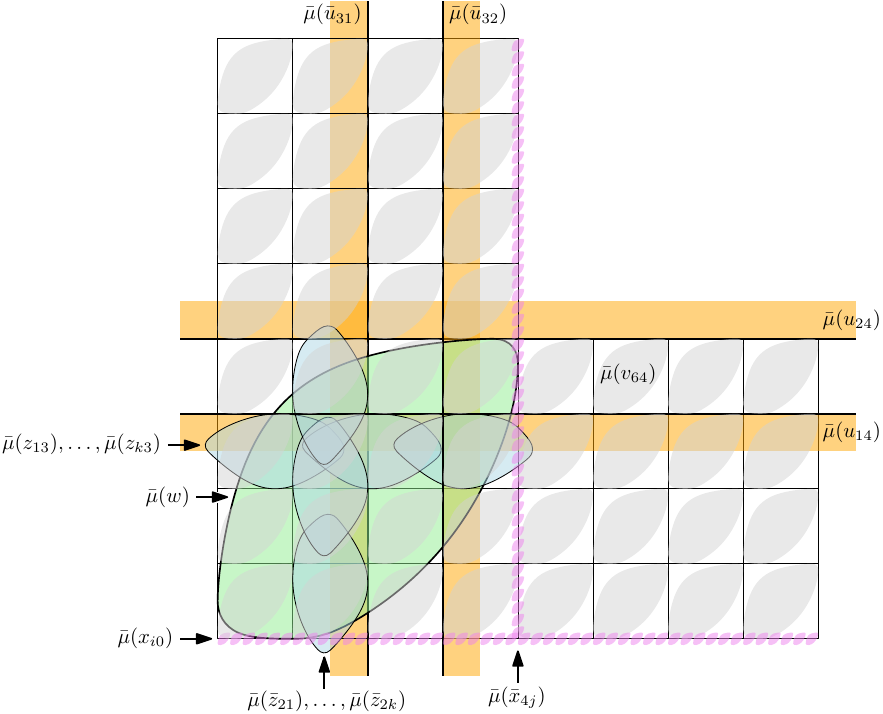}
\caption{The interior-realization of the graph $G_{4\mkern 1mu 8}^{A,\sim A}$.
The figure is not to scale; in reality, the pink disks $\intrep(x_{ij})$ and $\intrep(\bar x_{ij})$ would be much smaller (and thus more numerous).}
\label{fig:main construction}
\end{figure}

An \emph{$n$-chain} aligned to parallel lines $\ell_1,\ell_2$ is an $n$-tuple $A_1,\ldots,A_n$ of convex disks all touching $\ell_1$ and $\ell_2$ and such that $A_i\cap A_{i+1}\neq\emptyset$ for all $i\in[n-1]$.
The \emph{relative length} of such an $n$-chain is the length of the orthogonal projection of $A_1\cup\cdots\cup A_n$ on $\ell_1$ (or $\ell_2$) divided by $\dist(\ell_1,\ell_2)$.
Such an $n$-chain is \emph{strict} if $\int A_i\cap\int A_{i+1}\neq\emptyset$ for all $i\in[n-1]$.
A \emph{horizontal} or \emph{vertical\/ $n$-chain} is an $n$-chain aligned to horizontal or vertical lines, respectively.

The following construction describes the key graph of our proof; see Figure~\ref{fig:main construction}.

\begin{construction}[the graph $G_{mn}^{A,\calF}$]
\label{construction}
Let $A$ be a smooth convex disk with bounding box $[0,1]^2$.
Let $\calF=\hom A$ or $\calF=\sim A$.
Let $m,n\in\setN$ with $m\leq n$.
Let $k\in\setN$ be minimal such that there exist a strict horizontal $k$-chain $Z_1,\ldots,Z_k$ and a strict vertical $k$-chain $\bar Z_1,\ldots,\bar Z_k$ in $\calF$ of relative length greater than $m$.
Let $Z_1,\ldots,Z_k$ be a strict horizontal $k$-chain with bounding box $[-\delta,m+\delta]\times[0,1]$ and $\bar Z_1,\ldots,\bar Z_k$ be a strict vertical $k$-chain with bounding box $[0,1]\times[-\delta,m+\delta]$, for a suitably small $\delta>0$.
Let $\epsilon$ be the minimum of the values of $\epsilon_0$ claimed by Lemma~\ref{lemma:smoothness} for each of $A,Z_1,\ldots,Z_k,\bar Z_1,\ldots,\bar Z_k$ and $n$.
The graph $G_{mn}^{A,\calF}$ has the following vertices and the following interior-realization $\intrep$ by members of $\calF\cup\calH$:
\begin{itemize}
\item $\intrep(v_{ij})=\frac{1}{m}A+(\frac{i-1}{m},\frac{j-1}{m})$ for $(i,j)\in([n]\times[m])\cup([m]\times[n])$,
\item $\intrep(u_{1j})=\setR\times(-\infty,\frac{j-1}{m}]$ for $j=1,\ldots,m+1$ and $\intrep(u_{2j})=\setR\times[\frac{j}{m},+\infty)$ for $j=0,\ldots,m$,
\item $\intrep(\bar u_{i1})=(-\infty,\frac{i-1}{m}]\times\setR$ for $i=1,\ldots,m+1$ and $\intrep(\bar u_{i2})=[\frac{i}{m},+\infty)\times\setR$ for $i=0,\ldots,m$,
\item $\intrep(z_{ij})=\frac{1}{m}Z_j+(0,\frac{j-1}{m})$ for $(i,j)\in[k]\times[m]$,
\item $\intrep(\bar z_{ij})=\frac{1}{m}\bar Z_i+(\frac{i-1}{m},0)$ for $(i,j)\in[m]\times[k]$,
\item $\intrep(w)=A$,
\item $\intrep(x_{ij})=\frac{\epsilon}{m}A+(\frac{\epsilon i}{m},\frac{j}{m}-\frac{\epsilon}{2m})$ for $i=0,\ldots,\lceil\frac{n}{\epsilon}\rceil-1$ and $j=0,\ldots,m$,
\item $\intrep(\bar x_{ij})=\frac{\epsilon}{m}A+(\frac{i}{m}-\frac{\epsilon}{2m},\frac{\epsilon j}{m})$ for $i=0,\ldots,m$ and $j=0,\ldots,\lceil\frac{n}{\epsilon}\rceil-1$.
\end{itemize}
\end{construction}

By Lemma~\ref{lemma:interior-realization}, $G_{mn}^{A,\calF}$ has a realization in $\calF$.
When considering a specific realization $\rep$ of $G_{mn}^{A,\calF}$ (possibly with a superscript), we write $V_{ij}$, $U_{ij}$, $\bar U_{ij}$, $Z_{ij}$, $\bar Z_{ij}$, and $W$ (with the same superscript) as shorthand for $\rep(v_{ij})$, $\rep(u_{ij})$, $\rep(\bar u_{ij})$, $\rep(z_{ij})$, $\rep(\bar z_{ij})$, and $\rep(w)$, respectively.

For $m\in\setN$ and $i,j\in[m]$, let $S_{ij}^m=[\frac{i-1}{m},\frac{i}{m}]\times[\frac{j-1}{m},\frac{j}{m}]$.

\begin{lemma}
\label{lemma:properties}
Let\/ $A,\calF,m,n,k$ be as in Construction~\ref{construction}.
The following hold for\/ $G_{mn}^{A,\calF}$:
\begin{enumerate}
\item\label{prop:glue}
The following pairs of vertices form the small side of some induced subgraphs isomorphic to\/ $K_{2,n}$:
\begin{itemize}
\item $v_{ij}$ and\/ $u$ for each\/ $u\in\set{u_{1j},u_{2j},\bar u_{i1},\bar u_{i2}}$, for all\/ $(i,j)\in([n]\times[m])\cup([m]\times[n])$,
\item $z_{ij}$ and\/ $u_{1j}$, as well as\/ $z_{ij}$ and\/ $u_{2j}$, for all\/ $(i,j)\in[k]\times[m]$,
\item $\bar z_{ij}$ and\/ $\bar u_{i1}$, as well as\/ $\bar z_{ij}$ and\/ $\bar u_{i2}$, for all\/ $(i,j)\in[m]\times[k]$,
\item $w$ and\/ $u$ for each\/ $u\in\set{u_{11},u_{2m},\bar u_{11},\bar u_{m2}}$.
\end{itemize}
\item\label{prop:Ln}
For every\/ $j\in[m]$, there is an induced subgraph isomorphic to\/ $L_n$ in which the vertices\/ $u_{1j},u_{2j},u_{1(j+1)},u_{2(j-1)},v_{1j},\ldots,v_{nj}$ play the roles of\/ $u_1,u_2,\hat u_1,\hat u_2,v_1,\ldots,v_n$, respectively.
For every\/ $i\in[m]$, there is an induced subgraph isomorphic to\/ $L_n$ in which the vertices\/ $\bar u_{i1},\bar u_{i2},\bar u_{(i+1)1},\bar u_{(i-1)2},v_{i1},\ldots,v_{in}$ play the roles of\/ $u_1,u_2,\hat u_1,\hat u_2,v_1,\ldots,v_n$, respectively.
\item\label{prop:chain}
For every\/ $j\in[m]$, the subgraph induced on\/ $v_{1j},\ldots,v_{mj},z_{1j},\ldots,z_{kj}$ is connected and contains a path\/ $z_{1j}\cdots z_{kj}$; for every\/ $i\in[m]$, the subgraph induced on\/ $v_{i1},\ldots,v_{im},\bar z_{i1},\ldots,\bar z_{ik}$ is connected and contains a path\/ $\bar z_{i1}\cdots\bar z_{ik}$.
\item\label{prop:stretch}
The vertices\/ $z_{11},\ldots,z_{1m}$ are adjacent to\/ $\bar u_{11}$, the vertices\/ $z_{k1},\ldots,z_{km}$ are adjacent to\/ $\bar u_{m2}$, the vertices\/ $\bar z_{11},\ldots,\bar z_{m1}$ are adjacent to\/ $u_{11}$, and the vertices\/ $\bar z_{1k},\ldots,\bar z_{mk}$ are adjacent to\/ $u_{2m}$.
\item\label{prop:shape}
The vertex\/ $w$ is adjacent to at least one of\/ $z_{1j},\ldots,z_{kj}$ for every\/ $j\in[m]$ and at least one of\/ $\bar z_{i1},\ldots,\bar z_{ik}$ for every\/ $i\in[m]$.
\item\label{prop:grid}
For all\/ $i,j\in[m]$, if\/ $S_{ij}^m\subseteq A$, then\/ $v_{ij}w$ is an edge, and if\/ $v_{ij}w$ is an edge, then\/ $S_{ij}^m\cap A\neq\emptyset$.
\end{enumerate}
\end{lemma}

\begin{proof}
Let $\intrep$ be as in Construction~\ref{construction}.
Let $\calX_j=\set{\intrep(x_{ij})\colon i=0,\ldots,\lceil\frac{n}{\epsilon}\rceil-1}$ for $j=0,\ldots,m$ and $\bar\calX_i=\set{\intrep(\bar x_{ij})\colon j=0,\ldots,\lceil\frac{n}{\epsilon}\rceil-1}$ for $i=0,\ldots,m$.
By Lemma~\ref{lemma:smoothness}, for $(i,j)\in[n]\times[m]$, the lengths of the segments $\intrep(v_{ij})\cap(\setR\times\set{\frac{j-1}{m}+\frac{\epsilon}{2m}})$ and $\intrep(v_{ij})\cap(\setR\times\set{\frac{j}{m}-\frac{\epsilon}{2m}})$ are at least $\frac{n\epsilon}{m}$, so the interiors of these segments intersect at least $n$ disks in $\calX_{j-1}$ and $\calX_j$, respectively.
This implies that $\int\intrep(v_{ij})$ intersects the interiors of at least $n$ disks in $\calX_{j-1}$ and of at least $n$ disks in $\calX_j$.
Analogously,
\begin{itemize}
\item for $(i,j)\in[m]\times[n]$, $\int\intrep(v_{ij})$ intersects the interiors of at least $n$ disks in $\bar\calX_{i-1}$ and of at~least $n$ disks in $\bar\calX_i$;
\item for $(i,j)\in[k]\times[m]$, $\int\intrep(z_{ij})$ intersects the interiors of at least $n$ disks in $\calX_{j-1}$ and of at~least $n$ disks in $\calX_j$;
\item for $(i,j)\in[m]\times[k]$, $\int\intrep(\bar z_{ij})$ intersects the interiors of at least $n$ disks in $\bar\calX_{i-1}$ and of at~least $n$ disks in $\bar\calX_i$;
\item $\int\intrep(w)$ intersects the interiors of at least $n$ disks in each of $\calX_0$, $\calX_m$, $\bar\calX_0$, and $\bar\calX_m$.
\end{itemize}
This yields property~\ref{prop:glue}.
Property~\ref{prop:Ln} now follows from property~\ref{prop:glue}.
Properties \ref{prop:chain}, \ref{prop:stretch}, and~\ref{prop:shape} are straightforward.
Property~\ref{prop:grid} follows from the fact that $\intrep(v_{ij})\subseteq S_{ij}^m$ for $i,j\in[m]$.
\end{proof}

\begin{figure}[t]
\centering
\includegraphics[page=7]{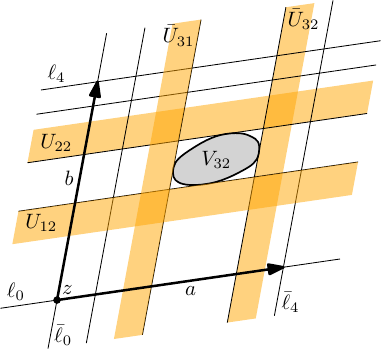}
\caption{An example of a $4$-grid with aligned disks and half-planes.}
\label{fig:grid}
\end{figure}

An \emph{$m$-grid} is a collection of two $(m+1)$-tuples of parallel lines $\ell_0,\ell_1,\ldots,\ell_m$ and $\bar\ell_0,\bar\ell_1,\ldots,\bar\ell_m$ that are images of horizontal lines at coordinates $0=y_0<y_1<\cdots<y_m=1$ and $m+1$ vertical lines at coordinates $0=x_0<x_1<\cdots<x_m=1$, respectively, under an affine transformation $f\colon\setR^2\ni(x,y)\mapsto z+xa+yb\in\setR^2$ for some point $z\in\setR^2$ called the \emph{origin} of the $m$-grid and some linearly independent vectors $a,b\in\setR^2$ that form the \emph{basis} of the $m$-grid; see Figure~\ref{fig:grid}.
The differences $x_1-x_0,\ldots,x_m-x_{m-1}$ and $y_1-y_0,\ldots,y_m-y_{m-1}$ are the \emph{horizontal} and \emph{vertical steps} of the $m$-grid, respectively.
A configuration of convex disks $V_{ij}$ with $i,j\in[m]$ and half-planes $U_{11},U_{21}\ldots,U_{1m},U_{2m},\bar U_{11},\bar U_{12},\ldots,\bar U_{m1},\bar U_{m2}$ is \emph{aligned} to such an $m$-grid if the following holds:
\begin{itemize}
\item $U_{1j}=f(\setR\times(-\infty,y_{j-1}])$ and $U_{2j}=f(\setR\times[y_j,+\infty))$ for $j\in[m]$,
\item $\bar U_{i1}=f((-\infty,x_{i-1}]\times\setR)$ and $\bar U_{i2}=f([x_i,+\infty)\times\setR)$ for $i\in[m]$,
\item $V_{ij}$ touches the four half-planes $U_{1j},U_{2j},\bar U_{i1},\bar U_{i2}$ for $i,j\in[m]$.
\end{itemize}

The following lemma is at the heart of our argument.
Among other things, it asserts that in realizations of the graph $G_{mn}^{A,\calF}$, the disks $V_{ij}^n$, for $i,j\in[m]$, are indeed forced to form an aligned $m$-grid as $n\to\infty$.
This will be the foundation for the proofs of Theorems \ref{thm:hom} and~\ref{thm:sim}.

\begin{lemma}
\label{lemma:grid}
Let\/ $A$ and\/ $B$ be smooth convex disks such that\/ $A$ has bounding box\/ $[0,1]^2$.
Let\/ $\calF=\hom A$ or\/ $\calF=\sim A$.
Let\/ $m\in\setN$.
Let\/ $k\in\setN$ be minimal such that there exist a strict horizontal\/ $k$-chain and a strict vertical\/ $k$-chain in\/ $\calF$ of length greater than\/ $m$.
Let\/ $(\rep^n)_{n\geq m}$ be a sequence of realizations\/ $\rep^n$ of\/ $G_{mn}^{A,\calF}$ in\/ $\sim B$ such that\/ $V_{11}^n$ is constant.
Then there is an infinite set\/ $N\subseteq\set{n\in\setN\colon n\geq m}$ on which
\begin{itemize}
\item the disks\/ $V_{ij}^n$ with\/ $i,j\in[m]$, $U_{1j}^n,U_{2j}^n$ with\/ $j\in[m]$, and\/ $\bar U_{i1}^n,\bar U_{i2}^n$ with\/ $i\in[m]$ converge to convex disks\/ $V_{ij}^\star\in\calF$ and half-planes\/ $U_{1j}^\star,U_{2j}^\star$ and\/ $\bar U_{i1}^\star,\bar U_{i2}^\star$, respectively, that are aligned to an\/ $m$-grid;
\item for every\/ $j\in[m]$, the disks\/ $Z_{1j}^n,\ldots,Z_{kj}^n$ converge to convex disks\/ $Z_{1j}^\star,\ldots,Z_{kj}^\star\in\calF$ each touching\/ $U_{1j}^\star$ and\/ $U_{2j}^\star$,
\item for every\/ $i\in[m]$, the disks\/ $\bar Z_{i1}^n,\ldots,\bar Z_{ik}^n$ converge to convex disks\/ $\bar Z_{i1}^\star,\ldots,\bar Z_{ik}^\star\in\calF$ each touching\/ $\bar U_{i1}^\star$ and\/ $\bar U_{i2}^\star$,
\item the disks\/ $W^n$ converge to a convex disk\/ $W^\star\in\calF$ touching\/ $U_{11}^\star$, $U_{2m}^\star$, $\bar U_{11}^\star$, and\/ $\bar U_{m2}^\star$.
\end{itemize}
\end{lemma}

\begin{proof}
Let $N=\set{n\in\setN\colon n\geq m}$.
In view of Lemma~\ref{lemma:properties} (\ref{prop:Ln} and~\ref{prop:chain}), we can apply Lemma~\ref{lemma:extended row} repeatedly as follows, with each application restricting the set $N$ to a suitable infinite subset:
\begin{itemize}
\item for each $j\in[m]$, with vertices $u_{1j},u_{2j},u_{1(j+1)},u_{2(j-1)}$, and $v_{1j},\ldots,v_{nj}$ playing the roles of $u_1,u_2,\hat u_1,\hat u_2$, and $v_1,\ldots,v_n$ (respectively) in $L_n$, and with the graph $H$ formed by $v_{1j},\ldots,v_{mj},z_{1j},\ldots,z_{kj}$,
\item for each $i\in[m]$, with vertices $\bar u_{i1},\bar u_{i2},\bar u_{(i+1)1},\bar u_{(i-1)2}$, and $v_{i1},\ldots,v_{in}$ playing the roles of $u_1,u_2,\hat u_1,\hat u_2$, and $v_1,\ldots,v_n$ (respectively) in $L_n$, and with the graph $H$ formed by $v_{i1},\ldots,v_{im},\bar z_{i1},\ldots,\bar z_{ik}$.
\end{itemize}
Specifically, we first use the assumption that $V_{11}^n$ is constant and proceed with the applications above for $j=1$ and $i=1$ to guarantee that $V_{21}^n,\ldots,V_{m1}^n$ and $V_{12}^n,\ldots,V_{1m}^n$ converge to convex disks, which is needed for the remaining applications above.
On the resulting restriction of $N$, the disks $V_{ij}^n$ with $i,j\in[m]$, $U_{1j}^n,U_{2j}^n,Z_{1j}^n,\ldots,Z_{kj}^n$ with $j\in[m]$, and $\bar U_{i1}^n,\bar U_{i2}^n,\bar Z_{i1}^n,\ldots,\bar Z_{ik}^n$ with $i\in[m]$ converge to limits $V_{ij}^\star$, $U_{1j}^\star,U_{2j}^\star,Z_{1j}^\star,\ldots,Z_{kj}^\star$, and $\bar U_{i1}^\star,\bar U_{i2}^\star,\bar Z_{i1}^\star,\ldots,\bar Z_{ik}^\star$, respectively, where
\begin{itemize}
\item $U_{1j}^\star$ and $U_{2j}^\star$ are disjoint half-planes for $j\in[m]$,
\item $U_{1(j+1)}^\star$ and $U_{2j}^\star$ touch and therefore share the boundary line, for $j\in[m-1]$,
\item $\bar U_{i1}^\star$ and $\bar U_{i2}^\star$ are disjoint half-planes for $i\in[m]$,
\item $\bar U_{(i+1)1}^\star$ and $\bar U_{i2}^\star$ touch and therefore share the boundary line, for $i\in[m-1]$,
\item $V_{ij}^\star$ is a convex disk touching $U_{1j}^\star$, $U_{2j}^\star$, $\bar U_{i1}^\star$, and $\bar U_{i2}^\star$ for $i,j\in[m]$,
\item $Z_{1j}^\star,\ldots,Z_{kj}^\star$ are convex disks each touching $U_{1j}^\star$ and $U_{2j}^\star$, for $j\in[m]$,
\item $\bar Z_{i1}^\star,\ldots,\bar Z_{ik}^\star$ are convex disks each touching $\bar U_{i1}^\star$ and $\bar U_{i2}^\star$, for $i\in[m]$.
\end{itemize}
Let
\begin{itemize}
\item $\ell_0=\partial U_{11}^\star$, $\ell_j=\partial U_{1(j+1)}^\star=\partial U_{2j}^\star$ for $j\in[m-1]$, and $\ell_m=\partial U_{2m}^\star$,
\item $\bar\ell_0=\partial\bar U_{11}^\star$, $\bar\ell_i=\partial\bar U_{(i+1)1}^\star=\partial\bar U_{i2}^\star$ for $i\in[m-1]$, and $\bar\ell_m=\partial\bar U_{m2}^\star$.
\end{itemize}
It follows that the lines $\ell_0,\ldots,\ell_m$ are parallel and occur in this order, and so do the lines $\bar\ell_0,\ldots,\bar\ell_m$.
Consequently, they form an $m$-grid, the origin of which is the intersection point of $\ell_0$ and $\bar\ell_0$, and the basis vectors of which are the vectors from the origin to the intersection point of $\ell_0$ and $\bar\ell_m$ and from the origin to the intersection point of $\ell_m$ and $\bar\ell_0$.
Furthermore, Lemma~\ref{lemma:distance} implies that $V_{ij}^\star$ touches $U_{1j}^\star,U_{2j}^\star,\bar U_{i1}^\star,\bar U_{i2}^\star$ for $i,j\in[m]$.
This shows that the disks $V_{ij}^\star$ with $i,j\in[m]$, $U_{1j}^\star,U_{2j}^\star$ with $j\in[m]$, and $\bar U_{i1}^\star,\bar U_{i2}^\star$ with $i\in[m]$ are aligned to the $m$-grid.

By Lemma~\ref{lemma:properties}~(\ref{prop:shape}), for every $n$, the vertex $w$ has an edge to at least one of the vertices $z_{ij}$ in $G_{mn}^{A,\calF}$ and therefore $W^n\cap Z_{ij}^n\neq\emptyset$.
It follows that the sequence $(W^n)_{n\in N}$ is anchored and therefore, restricting $N$ yet further by Lemma~\ref{lemma:limit}, $W^n$ converges to a limit $W^\star$.
By Lemma~\ref{lemma:properties}~(\ref{prop:shape}) and Lemma~\ref{lemma:distance}, $W^\star$ touches $U_{11}^\star,U_{2m}^\star,\bar U_{11}^\star,\bar U_{m2}^\star$.
Consequently, by Lemma~\ref{lemma:limit}, $W^\star$ is a convex disk.
\end{proof}

\section{Classifying Intersection Graphs of Homothets}\label{sec:homothets}

The proof of Theorem~\ref{thm:hom} is based on the following lemma.

\begin{lemma}
\label{lemma:homothets}
Let\/ $A$ and\/ $B$ be smooth convex disks such that\/ $A$ has bounding box\/ $[0,1]^2$.
If for all\/ $m,n\in\setN$ with\/ $m\leq n$, there is a realization of\/ $G_{mn}^{A,\hom A}$ in\/ $\hom B$, then there is an affine transformation that maps\/ $A$ to\/ $B$.
\end{lemma}

Before proving Lemma~\ref{lemma:homothets}, let us see how Theorem~\ref{thm:hom} follows.

\begin{proof}[Proof of Theorem~\ref{thm:hom}]
Let $A$ and $B$ be smooth convex disks.
As we already observed, if $A$ and $B$ are affine equivalent, then $\Ghom(A)=\Ghom(B)$, because the affine transformation that maps $A$ to $B$ transforms every realization in $\hom A$ to a realization of the same graph in $\hom B$, and vice versa.
Now, suppose $\Ghom(A)=\Ghom(B)$.
We can assume without loss of generality that the bounding box of $A$ is $[0,1]^2$, otherwise we can apply an affine transformation to $A$ to obtain a convex disk with that bounding box; as observed before, such a transformation does not change the intersection graphs realized in $\hom A$.
Now, since $G_{mn}^{A,\hom A}\in\Ghom(B)$ for all $m,n\in\setN$ with $m\leq n$, the lemma asserts that $A$ and $B$ are affine equivalent.

The last statement of the theorem asserts that when $A$ and $B$ are not affine equivalent, then the classes of intersection graphs are not nested.
Under this assumption, the lemma yields $G_{mn}^{A,\hom A}\notin\Ghom(B)$ for some $m$ and $n$.
Using the lemma with $A$ and $B$ interchanged, we also have $G_{mn}^{B,\hom B}\notin\Ghom(A)$ for some $m$ and $n$.
Therefore, the graph classes are not nested.
\end{proof}

\begin{proof}[Proof of Lemma~\ref{lemma:homothets}]
For all $m,n\in\setN$ with $m\leq n$, let $\rep^{mn}$ be a realization of $G_{mn}^{A,\hom A}$ in $\hom B$.
We first fix $m$ and consider the sequence of realizations $(\rep^{mn})_{n\geq m}$.
Without loss of generality, $V_{11}^{mn}$ is constant in this sequence.
By Lemma~\ref{lemma:grid}, we can pass to a subsequence such that the disks $V_{ij}^{mn}$ with $i,j\in[m]$, $U_{1j}^{mn},U_{2j}^{mn}$ with $j\in[m]$, and $\bar U_{i1}^{mn},\bar U_{i2}^{mn}$ with $i\in[m]$ converge to disks $V_{ij}^{m\star}\in\hom B$ and half-planes $U_{1j}^{m\star},U_{2j}^{m\star}$ and $\bar U_{i1}^{m\star},\bar U_{i2}^{m\star}$, respectively, that are aligned to an $m$-grid, and the disks $W^{mn}$ converge to a disk $W^{m\star}\in\hom B$.
It follows that all $V_{ij}^{m\star}$ with $i,j\in[m]$ have the same radius, so the horizontal and vertical steps of the $m$-grid are all equal to $\frac{1}{m}$.
Without loss of generality, the origin of the $m$-grid is $\origin$ and $r(W^{m\star})=1$.
Let $a^m,b^m\in\setR^2$ be the basis vectors of the $m$-grid, and let $f^m\colon\setR^2\ni(x,y)\mapsto xa^m+yb^m\in\setR^2$.
It follows that $V_{ij}^{m\star}\subseteq f^m(S_{ij}^m)$ for all $i,j\in[m]$ and $W^{m\star}\subseteq f^m([0,1]^2)$.

\begin{subclaim}
\label{claim:hom bound}
There is a constant $\eta>0$ such that $\norm{a^m}+\norm{b^m}\leq\eta$ for all $m$.
\end{subclaim}

\begin{claimproof}
There are at least $\lceil m^2\area A\rceil$ pairs $(i,j)\in[m]^2$ such that $S_{ij}^m\cap A\neq\emptyset$ and at most $4m$ such that $\emptyset\neq S_{ij}^m\cap A\neq S_{ij}^m$, so at least $\lceil m^2\area A\rceil-4m$ pairs $(i,j)\in[m]^2$ are such that $S_{ij}^m\subseteq A$.
For these pairs, $v_{ij}w$ is an edge of $G_{mn}^{A,\hom A}$ and consequently $W^{m\star}$ intersects $f^m(S_{ij}^m)$.
Likewise, there are at most $4m$ pairs $(i,j)\in[m]^2$ such that $\emptyset\neq f^m(S_{ij}^m)\cap W^{m\star}\neq f^m(S_{ij}^m)$, so at least $\lceil m^2\area A\rceil-8m$ pairs $(i,j)\in[m]^2$ are such that $f^m(S_{ij}^m)\subseteq W^{m\star}$.
Let $P^m=f^m([0,1]^2)$.
Since $r(W^{m\star})=1$ and
\[\frac{\area f^m(S_{ij}^m)}{\area P^m}=\frac{\area S_{ij}^m}{\area{[0,1]^2}}=\frac{1}{m^2}{,}\shortskipafterdisplay\]
the above implies
\[\frac{\area B}{\area P^m}=\frac{\area W^{m\star}}{\area P^m}\geq\frac{\lceil m^2\area A\rceil-8m}{m^2}\geq\area A-\frac{8}{m}\geq\frac{\area A}{2}{,}\]
where the last inequality holds for $m\geq 16/\area A$.
Let $\delta$ be the minimum distance between two parallel lines enclosing $B$.
Since $W^{m\star}\subseteq P^m$ and $r(W^{m\star})=1$, we have $\area P^m\geq\delta\max\set{\norm{a^m},\norm{b^m}}$.
Consequently, when $m\geq 16/\area A$, we have
\[\norm{a^m}+\norm{b^m}\leq 2\max\set{\norm{a^m},\norm{b^m}}\leq\frac{2}{\delta}\cdot\area P^m\leq\frac{4}{\delta}\cdot\frac{\area B}{\area A}{.}\qedhere\]
\end{claimproof}

\begin{subclaim}
\label{claim:hom convergence}
For every $\epsilon>0$, if $m$ is sufficiently large, then $d_H(W^{m\star},f^m(A))\leq\epsilon$.
\end{subclaim}

\begin{claimproof}
Let $\epsilon>0$.
Let $\eta$ be a constant from Claim~\ref{claim:hom bound}.
It follows that $\diam f^m(S_{ij}^m)\leq\frac{1}{m}(\norm{a^m}+\norm{b^m})\leq\frac{\eta}{m}$ for all $m$ and $i,j\in[m]$.
By Lemmas \ref{lemma:sequence from limit} and~\ref{lemma:convergence}, there are convex disks $\hat A\subset\int A$ and $\hat B\subset\int B$ such that $d_H(\hat A,A)\leq\smash[b]{\frac{\epsilon}{2\eta}}$ and $d_H(\hat B,B)\leq\frac{\epsilon}{2}$.
Let $\sigma=\dist(\hat A,\partial A)$ and $\tau=\dist(\hat B,\partial B)$, so that $\sigma,\tau>0$.
We show that $d_H(W^{m\star},f^m(A))\leq\epsilon$ whenever
\[m\geq\max\set[\bigg]{\frac{2\eta}{\epsilon},\:\frac{2}{\sigma},\:\frac{\eta}{\tau}}{.}\]

Let $p\in f^m(A)$.
We have $d_H(f^m(\hat A),f^m(A))\leq\eta d_H(\hat A,A)\leq\frac{\epsilon}{2}$, as $f^m$ stretches every segment by a factor of at most $\norm{a^m}+\norm{b^m}\leq\eta$.
Therefore, there is $\hat p\in f^m(\hat A)$ such that $\norm{p-\hat p}\leq\frac{\epsilon}{2}$.
Let $i,j\in[m]$ be such that $(f^m)^{-1}(\hat p)\in S_{ij}^m$.
It follows that $S_{ij}^m\subseteq\disk{(f^m)^{-1}(\hat p)}{2/m}\subseteq\disk{(f^m)^{-1}(\hat p)}{\sigma}\subseteq A$.
Therefore, by Lemma~\ref{lemma:properties}~(\ref{prop:grid}), $v_{ij}w$ is an edge of $G_{mn}^{A,\hom A}$, so there is a point $p'\in V_{ij}^{m\star}\cap W^{m\star}\subseteq f^m(S_{ij}^m)\cap W^{m\star}$.
It follows that $\norm{\hat p-p'}\leq\diam f^m(S_{ij}^m)\leq\frac{\eta}{m}\leq\frac{\epsilon}{2}$ and therefore $\norm{p-p'}\leq\norm{p-\hat p}+\norm{\hat p-p'}\leq\epsilon$.

Now, let $p\in W^{m\star}$.
Since $r(W^{m\star})=1$, there is a convex disk $\hat W\subset\int W^{m\star}$ such that $d_H(\hat W,W^{m\star})=d_H(\hat B,B)\leq\frac{\epsilon}{2}$ and $\dist(\hat W,\partial W^{m\star})=\dist(\hat B,\partial B)=\tau$.
By the former, there is $\hat p\in\hat W$ such that $\norm{p-\hat p}\leq\frac{\epsilon}{2}$.
Let $i,j\in[m]$ be such that $\hat p\in f^m(S_{ij}^m)$.
It follows that $V_{ij}^{m\star}\subseteq f^m(S_{ij}^m)\subseteq\disk{\hat p}{\eta/m}\subseteq\disk{\hat p}{\tau}\subseteq W^{m\star}$, so $v_{ij}w$ is an edge of $G_{mn}^{A,\hom A}$ and therefore, by Lemma~\ref{lemma:properties}~(\ref{prop:grid}), there is a point $p'\in f^m(S_{ij}^m)\cap f^m(A)$.
It follows that $\norm{\hat p-p'}\leq\frac{\epsilon}{2}$ (as before) and therefore $\norm{p-p'}\leq\norm{p-\hat p}+\norm{\hat p-p'}\leq\epsilon$.
\end{claimproof}

Since $\norm{a^m}+\norm{b^m}\leq\eta$ (by Claim~\ref{claim:hom bound}), there is an infinite set $M$ of indices $m$ such that $a^m$ and $b^m$ converge to vectors $a^\star,b^\star\in\setR^2$, respectively, on $M$.
Let $f^\star\colon\setR^2\ni(x,y)\mapsto xa^\star+yb^\star\in\setR^2$.
We show that $d_H(W^{m\star},f^\star(A))\to 0$ on $M$.
To this end, let $\epsilon>0$, and let $m\in M$ be sufficiently large that $d_H(W^{m\star},f^m(A))\leq\frac{\epsilon}{2}$ (by Claim~\ref{claim:hom convergence}) and $\norm{a^m-a^\star}+\norm{b^m-b^\star}\leq\frac{\epsilon}{2}$.
Since $A\subseteq[0,1]^2$, we have $\norm{f^m(x,y)-f^\star(x,y)}=\norm{x(a^m-a^\star)+y(b^m-b^\star)}\leq\norm{a^m-a^\star}+\norm{b^m-b^\star}\leq\frac{\epsilon}{2}$ for every point $(x,y)\in A$, whence it follows that $d_H(f^m(A),f^\star(A))\leq\frac{\epsilon}{2}$.
This yields $d_H(W^{m\star},f^\star(A))\leq d_H(W^{m\star},f^m(A))+d_H(f^m(A),f^\star(A))\leq\epsilon$.
By Lemma~\ref{lemma:convergence}, $W^{m\star}\to f^\star(A)$, so Lemma~\ref{lemma:limit} yields $f^\star(A)\in\hom B$, that is, there is a homothetic transformation $h\colon\setR^2\to\setR^2$ that maps $B$ to $f^\star(A)$.
We conclude that $h^{-1}\circ f^\star$ is an affine transformation that maps $A$ to $B$.
\end{proof}

\section{Classifying Intersection Graphs of Similarities}\label{sec:similarities}

For a convex disk $A$ and $n\in\setN$, we define $\sigma_A(n)$ as the maximum length of an $n$-chain in $\sim A$.
The sequence $(\sigma_A(n))_{n\in\setN}$ is subadditive, that is, $\sigma_A(n_1+n_2)\leq\sigma_A(n_1)+\sigma_A(n_2)$ for all $n_1,n_2\in\setN$.
Indeed, in an $(n_1+n_2)$-chain realizing the value $\sigma_A(n_1+n_2)$, the first $n_1$ disks form an $n_1$-chain of length $x_1\leq\sigma_A(n_1)$, and the last $n_2$ disks form an $n_2$-chain of length $x_2\leq\sigma_A(n_2)$, whence it follows that $\sigma_A(n_1+n_2)\leq x_1+x_2\leq\sigma_A(n_1)+\sigma_A(n_2)$.
By Fekete's Subadditive Lemma~\cite{fekete1923verteilung}, the limit $\lim_{n\to\infty}\sigma_A(n)/n$ exists and is equal to $\inf_{n\in\setN}\sigma_A(n)/n$.
We call this limit the \emph{stretch} of $A$ and denote it by $\rho_A$.

\begin{lemma}
\label{lemma:stretch lower bound}
For every\/ $k\in\setN$, we have\/ $\rho_Ak\leq\sigma_A(k)\leq\rho_Ak+\sigma_A(1)$.
\end{lemma}

\begin{proof}
The first inequality follows from the fact that $\rho_A=\inf_{n\in\setN}\sigma_A(n)/n$.
To see the second inequality, suppose for the sake of contradiction that there is $k\in\setN$ with $\sigma_A(k)>\rho_Ak+\sigma_A(1)$.
Let $A_1,\ldots,A_k$ be a $k$-chain of members of $\sim A$ of length $\sigma_A(k)$ aligned to horizontal lines at distance $1$.
Let $x$ be the maximum diameter of the intersection of $A_1\cup\cdots\cup A_k$ with a horizontal line.
It follows that $x\geq\sigma_A(k)-\sigma_A(1)>\rho_Ak$ and the two disks $A_k$ and $A_1+(x,0)$ touch in this order from left to right.
Consequently, for every $n\in\setN$, concatenating the $n$ translated copies
\[A_1+(jx,0),\:\ldots,\:A_k+(jx,0){,}\qquad\text{for }j=0,\ldots,n-1{,}\]
of the entire $k$-chain yields an $nk$-chain of length $\sigma_A(k)+(n-1)x$, showing that $\sigma_A(nk)\geq\sigma_A(k)+(n-1)x\geq nx$.
This leads to the following contradiction:
\[\rho_A=\lim_{n\to\infty}\frac{\sigma_A(n)}{n}=\lim_{n\to\infty}\frac{\sigma_A(nk)}{nk}\geq\frac{x}{k}>\rho_A{.}\qedhere\]
\end{proof}

Recall that $\simrefl A=\sim A\cup\sim\refl{A}$, where $\refl{A}$ is the reflection of $A$ about the $y$-axis.
The proof of Theorem~\ref{thm:sim} is based on the following lemma.

\begin{lemma}
\label{lemma:similarities}
Let\/ $A$ and\/ $B$ be smooth convex disks such that\/ $A$ has bounding box\/ $[0,1]^2$ and\/ $\rho_A\geq\rho_B$.
If for all\/ $m,n\in\setN$ with\/ $m\leq n$, there is a realization of\/ $G_{mn}^{A,\sim A}$ in\/ $\sim B$, then\/ $B\in\simrefl A$.
\end{lemma}

Before proving the lemma, let us see how Theorem~\ref{thm:sim} follows.

\begin{proof}[Proof of Theorem~\ref{thm:sim}]
Let $A$ and $B$ be smooth convex disks.
As we have already observed, if $B$ is similar to $A$ or to $\refl{A}$, then $\Gsim(A)=\Gsim(B)$, because the similarity transformation (possibly with reflection) that maps $A$ to $B$ transforms every realization in $\sim A$ to a realization of the same graph in $\sim B$, and vice versa.
Now, suppose $\Gsim(A)=\Gsim(B)$.
We can assume without loss of generality that $\rho_A\geq\rho_B$.
We can further assume that the bounding box of $A$ is $[0,1]^2$, otherwise we can rotate, scale, and translate $A$ to obtain a disk with this bounding box, and that transformation does not change the intersection graphs realized in $\sim A$.
Since $G_{mn}^{A,\sim A}\in\Gsim(B)$ for all $m,n\in\setN$ with $m\leq n$, the lemma yields $B\in\simrefl A$, as claimed.
\end{proof}

\begin{proof}[Proof of Lemma~\ref{lemma:similarities}]
For all $m,n\in\setN$ with $m\leq n$, let $\rep^{mn}$ be a realization of $G_{mn}^{A,\sim A}$ in $\sim B$.
We first fix $m$ and consider the sequence of realizations $(\rep^{mn})_{n\geq m}$.
Without loss of generality, $V_{11}^{mn}$ is constant in this sequence.
By Lemma~\ref{lemma:grid}, we can pass to a subsequence such that the disks $V_{ij}^{mn}$ with $i,j\in[m]$, $U_{1j}^{mn},U_{2j}^{mn}$ with $j\in[m]$, and $\bar U_{i1}^{mn},\bar U_{i2}^{mn}$ with $i\in[m]$ converge to disks $V_{ij}^{m\star}\in\sim B$ and half-planes $U_{1j}^{m\star},U_{2j}^{m\star}$ and $\bar U_{i1}^{m\star},\bar U_{i2}^{m\star}$, respectively, that are aligned to an $m$-grid, the disks $Z_{ij}^{mn}$ and $\bar Z_{ij}^{mn}$ converge to disks $Z_{ij}^{m\star}\in\sim B$ and $\bar Z_{ij}^{m\star}\in\sim B$, respectively, and the disks $W^{mn}$ converge to a disk $W^{m\star}\in\sim B$ that touches $U_{11}^\star,U_{2m}^\star,\bar U_{11}^\star,\bar U_{m2}^\star$.
Without loss of generality, the origin of the $m$-grid is $\origin$ and $r(W^{m\star})=1$.
Let $a^m,b^m\in\setR^2$ be the basis vectors of the $m$-grid, and let $f^m\colon\setR^2\ni(x,y)\mapsto xa^m+yb^m\in\setR^2$.
Let $\alpha_1^m,\ldots,\alpha_m^m$ and $\beta_1^m,\ldots,\beta_m^m$ be the horizontal and vertical steps of the $m$-grid, respectively, where $\sum_{i\in[m]}\alpha_i^m=\sum_{j\in[m]}\beta_j^m=1$.

\begin{figure}[t]
\centering
\includegraphics[page=8]{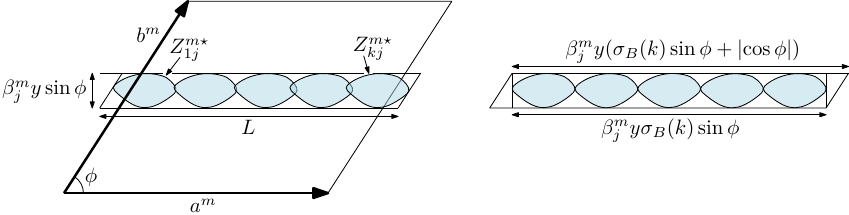}
\caption{To the left is shown the definition of the length $L$.
To the right is shown a maximum $k$-chain between two lines of distance $\beta_j^my\sin\phi$.
We have $\norm{a^m}\leq L\leq\beta_j^my(\sigma_B(k)\sin\phi+\abs{\cos\phi})$.}
\label{fig:claim}
\end{figure}

\begin{subclaim}
\label{claim:sim stretch}
There is a constant $c>0$ (which depends only on $B$) such that for every $m$, if $x=\norm{a^m}$, $y=\norm{b^m}$, and $\phi\in(0,\pi)$ is the angle between $a^m$ and $b^m$, then
\begin{gather*}
\tfrac{x}{y}\leq 1+\tfrac{c}{m}{,}\qquad\tfrac{y}{x}\leq 1+\tfrac{c}{m}{,}\qquad\sin\phi\geq 1-\tfrac{c}{m}{,}\\
\begin{alignedat}{5}
\tfrac{i}{m}-\tfrac{2c}{m}&<{}&\alpha_1^m&+\cdots+{}&\alpha_i^m&<{}&\tfrac{i}{m}+\tfrac{2c}{m}&\quad\text{for every }&i&\in[m-1]{,}\\
\tfrac{j}{m}-\tfrac{2c}{m}&<{}&\beta_1^m&+\cdots+{}&\beta_j^m&<{}&\tfrac{j}{m}+\tfrac{2c}{m}&\quad\text{for every }&j&\in[m-1]{.}
\end{alignedat}
\end{gather*}
\end{subclaim}

\begin{claimproof}
Let $x=\norm{a^m}$ and $y=\norm{b^m}$.
Let $\alpha_{\min}=\min_{i\in[m]}\alpha_i^m$ and $\beta_{\min}=\min_{j\in[m]}\beta_j^m$.
Let $\phi\in(0,\pi)$ be the angle between $a^m$ and $b^m$.
Let $k\in\setN$ be minimal such that there is a (strict) $k$-chain in $\sim A$ of length greater than $m$.
For each $j\in[m]$, the $k$-chain $Z_{1j}^{m\star},\ldots,Z_{kj}^{m\star}$ has length at most $\sigma_B(k)$ and is aligned to lines at distance $\beta_j^my\sin\phi$, so the length of the orthogonal projection of $Z_{1j}^{m\star}\cup\cdots\cup Z_{kj}^{m\star}$ on these lines is at most $\beta_j^my\sigma_B(k)\sin\phi$; see Figure~\ref{fig:claim}.
Consider the pair of lines parallel to $b^m$ and of minimum distance such that the entire chain $Z_{1j}^{m\star},\ldots,Z_{kj}^{m\star}$ is contained in the strip bounded by the lines, and let $L$ be the distance between the lines in the direction $a^m$.
We then get $L\leq\beta_j^my(\sigma_B(k)\sin\phi+\abs{\cos\phi})$, while we also have $L\geq x$, as $Z_{1j}^{m\star}$ intersects $\bar U_{11}^\star$ and $Z_{kj}^{m\star}$ intersects $\bar U_{m2}^\star$, by Lemma~\ref{lemma:properties}~(\ref{prop:stretch}).
This yields the following, using Lemma~\ref{lemma:stretch lower bound}, the inequality $\sigma_A(k-1)\leq m$ implied by the choice of $k$, and the assumption $\rho_A\geq\rho_B$:
\begin{align*}
x/(\beta_j^my)&\leq\sigma_B(k)\sin\phi+\abs{\cos\phi}\leq\rho_Bk\sin\phi+\sigma_B(1)\sin\phi+\abs{\cos\phi}\\
&\leq\rho_B(k-1)\sin\phi+\rho_B+\sigma_B(1)+1\leq\rho_A(k-1)\sin\phi+\rho_B+\sigma_B(1)+1\\
&\leq\sigma_A(k-1)\sin\phi+\rho_B+\sigma_B(1)+1\leq m\sin\phi+\rho_B+\sigma_B(1)+1.
\end{align*}
Thus $x\leq\beta_{\min}y(m\sin\phi+c)$ and analogously $y\leq\alpha_{\min}x(m\sin\phi+c)$, where $c=\rho_B+\sigma_B(1)+1$.
Since $\alpha_{\min},\beta_{\min}\leq\frac{1}{m}$, we have $x\leq y(\sin\phi+\frac{c}{m})$ and $y\leq x(\sin\phi+\frac{c}{m})$.
This yields $\frac{x}{y},\frac{y}{x}\leq\sin\phi+\frac{c}{m}\leq 1+\frac{c}{m}$ and $x+y\leq(x+y)(\sin\phi+\frac{c}{m})$, which yields $\sin\phi\geq 1-\frac{c}{m}$.
Moreover, $x\leq y(1+\frac{c}{m})\leq\alpha_{\min}xm(1+\frac{c}{m})^2$ yields $\alpha_{\min}\geq\frac{1}{m}(1+\frac{c}{m})^{-2}>\frac{1}{m}(1-\frac{2c}{m})$, and therefore, for every $i\in[m-1]$, we have
\[\tfrac{i}{m}-\tfrac{2c}{m}<\tfrac{i}{m}-\tfrac{i}{m}\cdot\tfrac{2c}{m}<i\alpha_{\min}\leq\alpha_1^m+\cdots+\alpha_i^m\leq 1-(m-i)\alpha_{\min}<\tfrac{i}{m}+\tfrac{m-i}{m}\cdot\tfrac{2c}{m}<\tfrac{i}{m}+\tfrac{2c}{m}{.}\]
Analogously, we have $\tfrac{j}{m}-\tfrac{2c}{m}<\beta_1^m+\cdots+\beta_j^m<\tfrac{j}{m}+\tfrac{2c}{m}$ for every $j\in[m-1]$.
\end{claimproof}

\begin{subclaim}
\label{claim:sim bound}
There is a constant $\eta>0$ such that $\norm{a^m}+\norm{b^m}\leq\eta$ for all $m$.
\end{subclaim}

\begin{claimproof}
By Claim~\ref{claim:sim stretch}, there is a constant $c$ such that for all $m$, if $\phi\in(0,\pi)$ is the angle between $a^m$ and $b^m$, then $\sin\phi\geq 1-\frac{c}{m}$.
The maximum distance between two parallel lines enclosing $B$ is at most $\diam B$, and it is at least $\norm{a^m}\sin\phi$ and at least $\norm{b^m}\sin\phi$, as $r(W^{m\star})=1$ and $W^{m\star}$ touches all four sides of the parallelogram $f^m([0,1]^2)$.
It follows that $\norm{a^m}+\norm{b^m}\leq 2\diam B/\sin\phi\leq 2\diam B/(1-\frac{c}{m})$.
\end{claimproof}

\begin{subclaim}
\label{claim:sim convergence}
For every $\epsilon>0$, if $m$ is sufficiently large, then $d_H(W^{m\star},f^m(A))\leq\epsilon$.
\end{subclaim}

\begin{claimproof}
Let $\epsilon>0$.
Let $\eta$ be a constant from Claim~\ref{claim:sim bound}.
For all $m$ and $i,j\in[m]$, let
\[\hat S_{ij}^m=[\alpha_1^m+\cdots+\alpha_{i-1}^m,\:\alpha_1^m+\cdots+\alpha_i^m]\times[\beta_1^m+\cdots+\beta_{j-1}^m,\:\beta_1^m+\cdots+\beta_j^m]{.}\]
It follows that $V_{ij}^{m\star}\subseteq f^m(\hat S_{ij}^m)$.
By Claim~\ref{claim:sim stretch}, the coordinates of the four sides of $\hat S_{ij}^m$ differ from the coordinates of the corresponding four sides of $S_{ij}^m$ by at most $\frac{2c}{m}$, which implies that $\diam f^m(S_{ij}^m\cup\hat S_{ij}^m)\leq\frac{4c+1}{m}(\norm{a^m}+\norm{b^m})\leq\frac{4c+1}{m}\eta$.
By Lemmas \ref{lemma:sequence from limit} and~\ref{lemma:convergence}, there are convex disks $\hat A\subset\int A$ and $\hat B\subset\int B$ such that $d_H(\hat A,A)\leq\smash[b]{\frac{\epsilon}{2\eta}}$ and $d_H(\hat B,B)\leq\frac{\epsilon}{2}$.
Let $\sigma=\dist(\hat A,\partial A)$ and $\tau=\dist(\hat B,\partial B)$, so that $\sigma,\tau>0$.
We show that $d_H(W^{m\star},f^m(A))\leq\epsilon$ whenever
\[m\geq\max\set[\bigg]{\frac{2(4c+1)\eta}{\epsilon},\:\frac{2}{\sigma},\:\frac{(4c+1)\eta}{\tau}}{.}\]

Let $p\in f^m(A)$.
We have $d_H(f^m(\hat A),f^m(A))\leq\eta d_H(\hat A,A)\leq\frac{\epsilon}{2}$, as $f^m$ stretches every segment by a factor of at most $\norm{a^m}+\norm{b^m}\leq\eta$.
Therefore, there is $\hat p\in f^m(\hat A)$ such that $\norm{p-\hat p}\leq\frac{\epsilon}{2}$.
Let $i,j\in[m]$ be such that $(f^m)^{-1}(\hat p)\in S_{ij}^m$.
It follows that $S_{ij}^m\subseteq\disk{(f^m)^{-1}(\hat p)}{2/m}\subseteq\disk{(f^m)^{-1}(\hat p)}{\sigma}\subseteq A$.
Therefore, by Lemma~\ref{lemma:properties}~(\ref{prop:grid}), $v_{ij}w$ is an edge of $G_{mn}^{A,\sim A}$, so there is a point $p'\in V_{ij}^{m\star}\cap W^{m\star}\subseteq f^m(\hat S_{ij}^m)\cap W^{m\star}$.
It follows that $\norm{\hat p-p'}\leq\diam f^m(S_{ij}^m\cup\hat S_{ij}^m)\leq\frac{4c+1}{m}\eta\leq\frac{\epsilon}{2}$ and therefore $\norm{p-p'}\leq\norm{p-\hat p}+\norm{\hat p-p'}\leq\epsilon$.

Now, let $p\in W^{m\star}$.
Since $r(W^{m\star})=1$, there is a convex disk $\hat W\subset\int W^{m\star}$ such that $d_H(\hat W,W^{m\star})=d_H(\hat B,B)\leq\frac{\epsilon}{2}$ and $\dist(\hat W,\partial W^{m\star})=\dist(\hat B,\partial B)=\tau$.
By the former, there is $\hat p\in\hat W$ such that $\norm{p-\hat p}\leq\frac{\epsilon}{2}$.
Let $i,j\in[m]$ be such that $\hat p\in f^m(\hat S_{ij}^m)$.
It follows that $V_{ij}^{m\star}\subseteq f^m(\hat S_{ij}^m)\subseteq\disk{\hat p}{(4c+1)\eta/m}\subseteq\disk{\hat p}{\tau}\subseteq W^{m\star}$, so $v_{ij}w$ is an edge of $G_{mn}^{A,\sim A}$ and therefore, by Lemma~\ref{lemma:properties}~(\ref{prop:grid}), there is a point $p'\in f^m(S_{ij}^m)\cap f^m(A)$.
It follows that $\norm{\hat p-p'}\leq\frac{\epsilon}{2}$ (as before) and therefore $\norm{p-p'}\leq\norm{p-\hat p}+\norm{\hat p-p'}\leq\epsilon$.
\end{claimproof}

Since $\norm{a^m}+\norm{b^m}\leq\eta$ (by Claim~\ref{claim:sim bound}), there is an infinite set $M$ of indices $m$ such that $a^m$ and $b^m$ converge to vectors $a^\star,b^\star\in\setR^2$, respectively, on $M$.
Let $f^\star\colon\setR^2\ni(x,y)\mapsto xa^\star+yb^\star\in\setR^2$.
It follows from Claim~\ref{claim:sim stretch} that $\norm{a^\star}=\norm{b^\star}$ and the vectors $a^\star$ and $b^\star$ are orthogonal, so $f^\star$ is a similarity transformation or similarity transformation with reflection.
The same argument as in the proof of Lemma~\ref{lemma:homothets}, using Claim~\ref{claim:sim convergence}, shows that $W^{m\star}\to f^\star(A)$.
By Lemma~\ref{lemma:limit}, it follows that $f^\star(A)\in\sim B$, and we have $f^\star(A)\in\simrefl A$, so $B\in\simrefl A$.
\end{proof}

\section{Open Problems}\label{sec:open}

For our row construction to work, we need the disks to be smooth.
In particular, Lemmas \ref{lemma:limit}, \ref{lemma:row}, and~\ref{lemma:smoothness} do not hold if $A$ is not smooth.
Distinguishing the classes of intersection graphs for non-smooth convex disks remains an interesting question.

One may also consider the even larger class $\Gaff(A)$ of intersection graphs of disks that are affine equivalent to a convex disk $A$ and ask when $\Gaff(A)=\Gaff(B)$ for two convex disks $A$ and $B$.
Other classes that have so far not been investigated are the contact and intersection graphs that can be obtained from rotated translations of a disk $A$, i.e., with no scaling allowed.

\bibliography{bibliography}

\end{document}